\documentclass[aps,prb,twocolumn]{revtex4}
\usepackage{graphicx}
\usepackage{color}
\usepackage{amssymb}
\def\v#1{\mathbf{#1}}
\newcommand{\St}{\underline{S}}
\newcommand{\Sd}{S^{\dagger}}
\newcommand{\Std}{\underline{S}^{\dagger}}
\newcommand{\gt}{\tilde{\gamma}}
\newcommand{\Gt}{\tilde{\Gamma}}
\newcommand{\xt}{\tilde{x}}
\newcommand{\Xt}{\tilde{X}}
\newcommand{\R}{r}
\newcommand{\T}{t}
\newcommand{\A}{a}
\newcommand{\D}{\mathfrak{D}}
\newcommand{\Area}{{\cal A}}
\begin{document}
\title{Nonequilibrium Superconductivity near Spin Active Interfaces}
\author{Erhai Zhao, Tomas L\"ofwander, and J. A. Sauls}
\affiliation{Department of Physics \& Astronomy,
             Northwestern University, Evanston, IL 60208}
\begin{abstract}
The Riccati formulation of the quasiclassical theory of nonequilibrium
superconductors is developed for spin-dependent scattering near magnetic
interfaces. We derive boundary conditions for the Riccati distribution
functions at a spin-active interface. The boundary conditions are formulated
in terms of an interface S-matrix describing the reflection and transmission
of the normal-state conduction electrons by the interface. The S-matrix
describes the effects of spin-filtering and spin-mixing (spin-rotation) by a
ferromagnetic interface. The boundary conditions for the Riccati equations are
applicable to a wide range of nonequilibrium transport properties of hybrid
systems of superconducting and magnetic materials. As an application we
calculate the spin and charge conductance of a normal
metal-ferromagnet-superconductor (NFS) point contact; the spin-mixing angle
that parameterizes the S-matrix is determined from experimental measurements
of the peak in the sub-gap differential conductance of the NFS point contact.
We also use the new boundary conditions to derive the effects of spin-mixing on the
phase-modulated thermal conductance of a
superconducting-ferromagnetic-superconducting (SFS) point contact. For
high-transparency (metallic ferromagnet) ``$\pi$'' junctions, the phase
modulation of the thermal conductance is dramatically different from that of
non-magnetic, ``0'' junctions. For low-transparency (insulating
ferromagnet) SFS tunnel junctions with weak spin mixing resonant
transmission of quasiparticles with energies just above the gap edge leads to
an increase of the thermal conductance, compared to the normal-state
conductance at $T_c$, over a broad temperature range when the superconducting
phase bias is $\phi\gtrsim \pi/2$.
\end{abstract}
\maketitle

\section{Introduction}

Spin dependent transport in hybrid systems composed of superconductors and
spin-active materials such as ferromagnets has attracted a lot of attention
because of the possibility of generating coherent spin transport for
spintronic devices.\cite{wol01} The spin polarization of a
ferromagnetic material, one of the key parameters in the development of
spintronic devices, is usually measured either by spin-dependent tunnelling
techniques,\cite{ted94} or by point-contact Andreev reflection
spectroscopy.\cite{sou98} Both methods infer information about the spin
polarization from the conductance data of superconductor-ferromagnet (SF)
junctions. When sandwiched between two $s$-wave superconducting leads (an SFS
junction), a ferromagnetic layer can produce a ``$\pi$ junction'', i.e. a
ground state of an SFS junction in which there is a $\pi$ phase difference
between the two superconductors.\cite{buz82,buz91} The $\pi$ state has been
observed in SFS junctions with metallic ferromagnetic
layers.\cite{rya01,kon02,gui03} It is predicted theoretically that an
insulating or semiconducting ferromagnetic layer can also produce a $\pi$
junction.\cite{fog00,bar02c} More complicated $\pi$ junctions in which the
Josephson coupling is provided by an inhomogeneous magnetization, e.g. a
ferromagnet-insulator-ferromagnet (FIF) tri-layer, have also been investigated
theoretically.\cite{kos01,ber01,gol02,bar02b}

The study of SFS junctions is fuelled in part by the proposal that $\pi$
junctions can be used to construct a non-dissipative superconducting phase
qubit.\cite{bla01} Most theoretical investigations of SFS junctions are
restricted to equilibrium properties, however the performance of a $\pi$
junction as a qubit depends sensitively on the suppression of dissipative
dynamics under nonequilibrium conditions. Recently, the nonequilibrium transport
properties of Josephson junctions with spin active interfaces have begun to be
explored theoretically.\cite{and02}

A powerful formulism for calculating the nonequilibrium properties of
superconducting heterostructures is provided by the quasiclassical theory
of superconductivity.\cite{eli72,lar76,ser83,ram86} Traditionally one obtains
the quasiclassical Green's functions by solving the transport equations subject
to boundary conditions at surfaces or interfaces.

A multi-step approach to the boundary value problem at a surface or interface
based on an interface transition matrix has been used by several
authors.\cite{ser83,kie85,cue01,esc03,kop04} This method requires one to
calculate an auxiliary Green's function for an impenetrable surface. The auxiliary
Green's function is then used as an input to a T-matrix equation from which
one constructs the quasiclassical Green's functions at the interface. This method
can be applied to a broad class of interface models, and is suitable for numerical
computations,\cite{cue01,esc03,kop04} but it requires the computation of
intermediate, unphysical auxiliary Green's functions.

Boundary conditions which are expressed only in terms of the physical quasiclassical
propagators and interface reflection and transmission amplitudes have been derived
from microscopic scattering theory by Zaitsev\cite{zai84} and Kieselmann\cite{kie85}
for nonmagnetic interfaces, and for spin active interfaces by Millis, Rainer and one
of the authors.\cite{mil88} These boundary conditions are formulated as a set of
third order equations in terms of the matrix Green's functions at the boundary, connected
via an interface scattering matrix (S-matrix) for normal metal electrodes. Although
auxiliary propagators are not present, the nonlinear boundary conditions are
non-intuitive, difficult to solve and contain unphysical solutions which must be
discarded.

Recently, a more intuitive and computationally efficient form of the
quasiclassical boundary conditions was obtained for non-magnetic interfaces by
Eschrig.\cite{esc00} This formulation starts from the boundary condition of
Zaitsev and Kieselmann, and is obtained by parameterizing the quasiclassical
Green's functions in terms of Riccati
amplitudes.\cite{sch95,sch96,esc99,she00,esc00} By formulating the quasiclassical
theory in terms of the Riccati amplitudes, not only are the transport equations
easier to solve numerically, but the boundary conditions become linear and free of
spurious solutions.\cite{esc00} Eschrig's formulation of the boundary condition
amounts to finding physical solution to the Zaitsev-Kieselmann nonlinear boundary
condition. For spin-active interfaces Fogelstr\"om obtained boundary conditions
for the retarded and advanced coherence functions.\cite{fog00}
However, a complete set of boundary conditions for non-equilibrium transport with
spin-active interfaces was lacking.

In this paper we derive the boundary condition for the quasiclassical Riccati
amplitudes, both the coherence functions and distribution functions, for spin
active interfaces and apply the new boundary conditions to study the
nonequilibrium transport properties of clean superconductor-ferromagnet hybrid
systems. The paper is organized as follows. The complete set of boundary
conditions for the Riccati amplitudes at spin active interfaces is presented in
section II, with technical steps of the derivation described in an appendix. In
section III, the S-matrix for scattering by two models for spin-active
interfaces, a ferromagnetic-insulating layer and a ferromagnetic-metallic layer,
are derived and discussed in terms of the effects of \textsl{spin filtering} and
\textsl{spin mixing}. Applications of the theory to the conductance of
the normal metal-ferromagnet-superconductor junction is analyzed in section IV.
In section V the influence of spin mixing on the phase
sensitive heat transport in SFS point contact is discussed in detail.

\section{The boundary conditions for Riccati amplitudes\label{sec_boundary-conditions}}

In the Riccati formulation of the quasiclassical theory of nonequilibrium
superconductivity, the quasiparticle excitation spectrum is determined from coherence
functions, $\gamma^{R,A}$ and $\tilde{\gamma}^{R/A}$, which measure the relative
amplitudes for normal-state quasi-particle and quasi-hole excitations; the occupation
probability of theses states is described by distribution functions, $x^K$ and
$\tilde{x}^K$.\cite{sch95,sch96,esc99,she00,esc00} For brevity we refer to both types of
functions as Riccati amplitudes, or Riccati functions, since all obey
Riccati-type transport equations, defined on classical trajectories in phase space
($\v{p},\v{R}$). Thus, in general the Riccati amplitudes are functions of space,
$\v{R}$, time, $t$, the direction of the Fermi momentum, $\v{p}_f$ (or Fermi velocity
$\v{v}_f$) and the excitation energy, $\epsilon$.

The Riccati amplitudes depend on spin, and in general are described by
$2\times 2$ density matrices in spin space whose eigenvalues determine the local
coherence and distribution functions for two possible spin states.
The coherence amplitudes obey Riccati-type equations; for example,
\begin{equation}
i\v{v}_f\cdot\v{\nabla}
\gamma^R=-2\epsilon\gamma^R+\gamma^R\tilde{\Delta}\gamma^R+
          \Sigma^R\gamma^R-\gamma^R\tilde{\Sigma}^R-\Delta^R
\,.
\end{equation}
The distribution function, $x^K$, obeys a Riccati-type transport equation,
\begin{eqnarray}
i(\partial_t+\v{v}_f\cdot\v{\nabla})x^K  &=& (\gamma^R\tilde{\Delta}^R+\Sigma^R)x^K
\nonumber\\
&+& x^K(\Delta^A\tilde{\gamma}^A-\Sigma^A)
\nonumber\\
&-&\gamma^R\tilde{\Sigma}^K\tilde{\gamma}^A -\Sigma^K
\nonumber\\
&+&\Delta^K\tilde{\gamma}^A + \gamma^R\tilde{\Delta}^K
\,,
\end{eqnarray}
where $\Sigma^{\mu}$ and $\Delta^{\mu}$, $\mu=R,A,K$, are the diagonal and
off-diagonal self-energies, respectively. We follow the notation in Ref.
\onlinecite{esc00} throughout the paper. Particle$\leftrightarrow$hole
conjugation, denoted by $\tilde{ }$, is defined by the operation
$\tilde{q}(\hat{\v{p}}_f,\epsilon)=q^*(-\hat{\v{p}}_f,-\epsilon)$. The product of two
functions of energy and time is defined by the non-commutative convolution,
\begin{equation}
AB\equiv A
\circ
B(\epsilon,t)=e^{i(\partial^A_{\epsilon}\partial^B_t-
                   \partial^A_{t}\partial^B_{\epsilon})/2}A(\epsilon,t)
B(\epsilon,t)
\,.
\end{equation}
Neither the operator, $\circ$, nor the arguments, $(\epsilon,t)$, are
shown explicitly unless required.

Once the Riccati equations are solved subject to appropriate boundary conditions,
the quasiclassical Green's functions can be constructed from the Riccati
amplitudes. Physical observables such as the charge or heat current can then be
calculated. This procedure is discussed extensively by several authors, c.f. Refs.
\onlinecite{esc99,esc00}.

At an interface or surface the local electronic potential changes on an atomic
length and energy scales. Such strong, short-range potentials are treated within the
quasiclassical theory as boundary conditions for the quasiclassical Green's
functions, or equivalently the Riccati amplitudes. Such an interface can be
described by a scattering matrix, $\mathcal{S}$, for normal-state electrons and
holes with excitation energies near the Fermi surface.\cite{mil88} Here we confine
our discussion to specular interfaces, in which case the momentum of an excitation
parallel to the interface, $\v{p}_{||}$, is conserved. The interface $\mathcal{S}$
matrix is then described by a unitary matrix in the combined spin, particle-hole and
direction spaces. Thus,
\begin{equation}
\mathcal{S}(\hat{\v{p}}_f)=\left(
\begin{array}{cc}
\hat{S}_{11} & \hat{S}_{12} \\
\hat{S}_{21} & \hat{S}_{22}
\end{array}
\right)
\,,
\end{equation}
where the index 1 (2) refers to the left (right) side of the interface. Each element of
this matrix, $\hat{S}_{ij}$, is a diagonal Nambu matrix in particle-hole space,
\begin{equation}
\hat{S}_{ij}=\left(
\begin{array}{cc}
 S_{ij} & 0  \\
 0 & \underline{S}_{ij}
\end{array} \right),
\; i,j=1,2;
\end{equation}
in which $S_{ij}$ and $\underline{S}_{ij}$ are matrices in spin
space, which are related by particle-hole conjugation,
$\underline{S}_{ij}(\mathbf{p}_{\parallel})=[S_{ji}(-\mathbf{p}_{\parallel})]^{\text{tr}}$,
where $[..]^{\text{tr}}$ is the matrix transpose in spin space.\cite{mil88}
\begin{figure}
\includegraphics[width=3.3in]{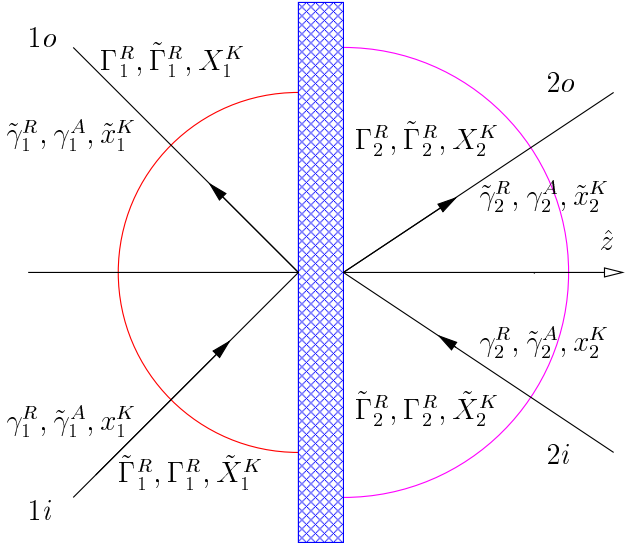}
\caption{The Riccati amplitudes corresponding to the scattering trajectories
for a partially transmitting specular interface. The interface normal is $\hat{z}$.
The trajectories for the scattering states are labelled $1i, 1o, 2i$ and $2o$.}
\label{fig1}
\end{figure}

The Riccati amplitudes for the set of scattering trajectories labelled $1i, 1o,
2i$ and $2o$, shown in Fig. \ref{fig1}, are classified into two groups. The
quantities $\{\gamma^{R,A}_j,\tilde{\gamma}^{R,A}_j,x^K_j,\tilde{x}^K_j\}$,
denoted by lower case symbols, are obtained by integrating the Riccati equations
along the four trajectories from the bulk toward the interface. The quantities
$\{\Gamma^{R,A}_j,\tilde{\Gamma}^{R,A}_j,X^K_j,\tilde{X}^K_j\}$, denoted by upper
case symbols, are obtained by starting at the interface and integrating along the
trajectory into the bulk. The boundary conditions listed below connect the unknown
upper case amplitudes at the interface with the known lower case amplitudes via
the interface scattering matrix. The derivation of these boundary conditions is
outlined in the appendix. For example, the boundary
conditions for trajectory $1o$ are,
\begin{eqnarray}
\Gamma^R_1 &=&\R^R_{1l}\gamma^R_1 \Std_{11} +
\T^R_{1l}\gamma^R_2 \Std_{12}  \label{first}\,,\\
\Gt^A_1&=&\St_{11}\gt^A_1\tilde{\R}^A_{1r}
+\St_{12}\gt^A_2\tilde{\T}^A_{1r} \,,\\
X^K_1 &=& \R^R_{1l} x^K_1 \tilde{\R}^A_{1r} + \T^R_{1l} x^K_2
\tilde{\T}^A_{1r} - \A^R_{1l} \xt^K_2 \tilde{\A}^A_{1r }
\,,
\label{third}
\end{eqnarray}
where we have introduced effective reflection ($\R$), transmission ($\T$)
and branch-conversion transmission ($\A$) amplitudes,
\begin{eqnarray}
\R^R_{1l}&=&+
[(\beta^R_{21})^{-1}\Sd_{11}-(\beta^R_{22})^{-1}\Sd_{12}]^{-1}
(\beta^R_{21})^{-1} \,, \\
\T^R_{1l}&=&-[(\beta^R_{21})^{-1}\Sd_{11}-(\beta^R_{22})^{-1}\Sd_{12}]^{-1}
(\beta^R_{22})^{-1}  \,, \\
\tilde{\R}^A_{1r}&=&+(\beta^A_{21})^{-1}
[S_{11}(\beta^A_{21})^{-1} - S_{12}(\beta^A_{22})^{-1}]^{-1} \,, \\
\tilde{\T}^A_{1r}&=&-(\beta^A_{22})^{-1}
[S_{11}(\beta^A_{21})^{-1} - S_{12}(\beta^A_{22})^{-1}]^{-1}  \,, \\
\A^R_{1l}&=&(\Gamma^R_1
\St_{11}-S_{11}\gamma^R_1)(\tilde{\beta}^R_{12})^{-1}  \,, \\
\tilde{\A}^A_{1r}
&=&(\tilde{\beta}^A_{12})^{-1}(\Std_{11}\Gt^A_1-\gt^A_1\Sd_{11})
\label{fourth}
\,.
\end{eqnarray}
The auxiliary quantities, $\beta^{R/A}_{ij}$, are defined as
\begin{eqnarray}
\beta^R_{ij}&=&\Sd_{ij}-\gamma^R_j\Std_{ij}\gt^R_i; \;
\tilde{\beta}^R_{ij}=\St_{ji}-\gt^R_j S_{ji} \gamma^R_i \,, \\
\beta^A_{ij}&=&S_{ij}-\gamma^A_i\St_{ij}\gt^A_j; \;
\tilde{\beta}^A_{ij}=\St^{\dagger}_{ji}-\gt^A_i \Sd_{ji}
\gamma^A_j
\,.
\end{eqnarray}
Similarly, for trajectory $1i$ we have
\begin{eqnarray}
\Gt^R_1&=& \tilde{\R}^R_{1l}\gt^R_1 S_{11} +
\tilde{\T}^R_{1l}\gt^R_2 S_{21} \,,\\
\Gamma^A_1&=&\Sd_{11}\gamma^A_1\R^A_{1r}
+\Sd_{21}\gamma^A_2\T^A_{1r} \,,\\
\Xt^K_1 &=& \tilde{\R}^R_{1l}\xt^K_1 \R^A_{1r} +
\tilde{\T}^R_{1l}\xt^K_2 \T^A_{1r} - \tilde{\A}^R_{1l} x^K_2
\A^A_{1r}
\,,
\label{seven}
\end{eqnarray}
where
\begin{eqnarray}
\tilde{\R}^R_{1l}&=&+[(\tilde{\beta}^R_{21})^{-1}\St_{11}-
                    (\tilde{\beta}^R_{22})^{-1}\St_{21}]^{-1}
                    (\tilde{\beta}^R_{21})^{-1}  \,,\\
\tilde{\T}^R_{1l}&=&-[(\tilde{\beta}^R_{22})^{-1}\St_{11}-
                    (\tilde{\beta}^R_{22})^{-1}\St_{21}]^{-1}
                    (\tilde{\beta}^R_{21})^{-1}  \,,\\
\R^A_{1r}&=&+(\tilde{\beta}^A_{21})^{-1}[\Std_{11}
              (\tilde{\beta}^A_{21})^{-1}-
               \Std_{21}(\tilde{\beta}^A_{22})^{-1}]^{-1}\,, \\
\T^A_{1r}&=& -(\tilde{\beta}^A_{22})^{-1}
[\Std_{11}(\tilde{\beta}^A_{21})^{-1}-\Std_{21}
          (\tilde{\beta}^A_{22})^{-1}]^{-1} \,, \\
\tilde{\A}^R_{1l}&=&(\Gt^R_1
\Sd_{11}-\Std_{11}\gt^R_1)(\beta^R_{12})^{-1}  \,, \\
\A^A_{1r}
&=&(\beta^A_{12})^{-1}(S_{11}\Gamma^A_1-\gamma^A_1\St_{11})
\,.
\label{final}
\end{eqnarray}
The boundary conditions for trajectories $2i$
and $2o$ are given by interchanging indices
$1\leftrightarrow 2$ in Eqs. (\ref{first})-(\ref{final}).
The derivation of Eqs. (\ref{first})-(\ref{final}) is described in
the appendix.

Note that there is more than one physically equivalent representation
of the boundary condition for any of the coherence functions. For example,
it is straightforward to show that an alternative form of the boundary condition
in Eq. \ref{first} for $\Gamma^R_1$ is given by
\begin{eqnarray}
\Gamma^R_1 &=& S_{11}\gamma^R_1 \R^R_{1r} + S_{12}\gamma^R_2\T^R_{1r}
\label{alternative}
\\
\R^R_{1r}&=&+(\tilde{\beta}^R_{12})^{-1}[\St_{11}(\tilde{\beta}^R_{12})^{-1}
-\St_{12}(\tilde{\beta}^R_{22})^{-1} ]^{-1}  \\
\T^R_{1r}&=&-
(\tilde{\beta}^R_{22})^{-1}[\St_{11}(\tilde{\beta}^R_{12})^{-1}
-\St_{12}(\tilde{\beta}^R_{22})^{-1} ]^{-1}
\,.
\end{eqnarray}
Similar results for $\Gamma^{R,A}_j$ and $\Gt^{R,A}_j$ were obtained by
Fogelstr\"om.\cite{fog00} Combined with these results for the coherence
functions, the boundary conditions for distribution functions given in Eqs.
(\ref{third}) and (\ref{seven}) provide a complete set of quasiclassical
boundary conditions applicable to a wide range of non-equilibrium conditions
for superconductors in contact with spin-active interfaces. These boundary
conditions (Eqs. (\ref{first})-(\ref{final})) reduce to the results of Ref.
\onlinecite{esc00} for non-spin-active scattering, i.e. when $S_{ij}$ and
$\St_{ij}$ are spin-independent.\footnote{
In comparing the Eqs. (\ref{first})-(\ref{final}) for spin-inactive interfaces
with the results of Ref. \onlinecite{esc00} note that $\R,\T,\A$ are effective
reflection (transmission) \textit{amplitudes}. For example, from Eq. (\ref{fourth})
it is seen that $\R_{1l}^R$ and $\T_{1l}^R$ satisfy
$\R_{1l}^R S_{11}^{\dagger}+\T_{1l}^R S_{12}^{\dagger}=1$ and
$\R_{1l}^R \beta_{21}^R+\T_{1l}^R \beta_{22}^{R}=0$.
These amplitudes are related to, but
should not be confused with the effective reflection (transmission)
\textit{probabilities},
$\mathcal{R},\mathcal{D},\mathcal{A}$ of Ref. \onlinecite{esc00}
which satisfy, $\mathcal{R}_{1l}^R +\mathcal{D}_{1l}^R =1$.
}

In deriving Eqs. (\ref{first})-(\ref{final}) we assumed that the
inverses of $\beta^{R/A}_{ij}$, and their $\tilde{\;}$ partners,
are defined. Equations (\ref{first})-(\ref{final}) cannot be
applied when one or more of the $\mathcal{S}$ matrix elements is zero.
However, in cases where this happens the boundary conditions are
significantly simplified, and can be readily derived following
the procedure outlined in the appendix. For example, in the case
of an impenetrable wall, we have perfect reflection
described by $S_{12}=S_{21}=0$.  Then Eqs. (\ref{first})-(\ref{third})
are replaced by the simpler set of boundary conditions,

\begin{eqnarray}
\Gamma^R_1&=&S_{11}\gamma^R_1\underline{S}_{11}^{\dagger} \\
\Gt^A_1&=&\underline{S}_{11}\gt^A_1 S_{11}^{\dagger} \\
X^K_1&=&S_{11}x^K_1 S_{11}^{\dagger}
\,.
\end{eqnarray}
Similarly for perfect transmission, $S_{jj}=0$, the boundary
conditions simplify to
\begin{eqnarray}
\Gamma^R_1&=&S_{12}\gamma^R_2\underline{S}_{12}^{\dagger} \\
\Gt^A_1&=&\underline{S}_{12}\gt^A_2 S_{12}^{\dagger} \\
X^K_1&=&S_{12}x^K_2S_{12}^{\dagger}
\,.
\end{eqnarray}

Further simplification occurs for stationary non-equilibrium transport.
The time convolution products reduce to matrix products, and there are
additional symmetry relations: $\beta^A_{ij}=(\beta^R_{ij})^{\dagger}$,
$\tilde{\beta}^A_{ij}=(\tilde{\beta}^R_{ij})^{\dagger}$,
$\tilde{\R}^A_{1r}=(\R^R_{1l})^{\dagger}$,
$\tilde{\T}^A_{1r}=(\T^R_{1l})^{\dagger}$, and
$\tilde{\A}^A_{1r}=(\A^R_{1l})^{\dagger}$.

\section{The $\mathcal{S}$ matrix\label{sec_Smatrix}}

A microscopic calculation of the normal-state $\mathcal{S}$ matrix for a spin active
interface would require a solution of the many-body problem in the presence of the
interface potential. This is a formidable problem and outside the realm of a practical
theory aimed at understanding the transport properties of heterogeneous superconducting
junctions. The alternative approach is to identify the structure of the $\mathcal{S}$
matrix, including the constraints of symmetry, and then model the interface in terms of
the key physical parameters defining these characteristics, e.g. the transmission and
reflection probabilities for normal-state electrons and holes moving along specific
trajectories and in particular spin states. For a relatively small set of physical
parameters, the key characteristics of the interface can be obtained from measurements,
e.g. from normal-state transport properties, and then used to make predictions for
non-equilibrium properties in the superconducting state. This is the most tractable
approach to interpreting and predicting the nonequilibrium properties of heterogeneous
superconducting junctions.

In this section we discuss the parametrization of the $\mathcal{S}$ matrix in terms of a
spin-mixing angle and spin-dependent normal-state transmission coefficients. Other
authors have also discussed the form of this interface $\mathcal{S}$ matrix for
particular magnetic interfaces, c.f. Refs. \onlinecite{tok88,fog00,bar02c,bar02b} We
discuss the form of the $\mathcal{S}$ matrix for both a ferromagnetic insulating
interface and a clean ferromagnetic metallic interface. For both cases we assume the
interface is atomically smooth so that the momentum parallel to the interface,
$\v{p}_{||}$, is a good quantum number.

First consider the $\mathcal{S}$ matrix of a ferromagnetic insulating or semiconducting
(FI) interface.\cite{tok88,fog00,bar02c} Choose the direction of the spontaneous
magnetization $\hat{\mu}$ as the quantization axis for the conduction electron spin.
Then spin up $(+)$ and down $(-)$ electrons see the FI interfaces as a potential barrier
with thickness $l$ and height $E_g\mp h$, where $E_g$ is the average band gap and $h$ is
the exchange energy. The effects of the FI layer on the transport of electrons are two
fold: 1) \emph{spin filtering} in which the reflection (transmission)
probabilities for spin up and spin down electrons are different,
because these electrons with different spin polarization see different potential barriers,
and 2) \emph{spin mixing} in which spin up and down electrons acquire different phase shifts
upon reflection (transmission). This is the analog of circular birefringence in optics.
Thus, in general the polarization of an incident electron undergoes a rotation analogous
to optical Faraday rotation.

The degrees of spin filtering and spin mixing are determined by the modulus and
the phase of the reflection (transmission) amplitudes, respectively. Consider the
reflection amplitude for example. In the spinor basis $|\pm\rangle$ which
diagonalizes $\hat{\mu}\cdot\vec{\sigma}$, the spin matrix $S_{11}$ is diagonal,
\begin{equation}
S_{11}=\left(
\begin{array}{cc}
  r_+ & 0 \\
  0 & r_- \\
\end{array}\right)=
\left(
\begin{array}{cc}
  |r_+|e^{i\theta_+} & 0 \\
  0 & |r_-|e^{i\theta_-} \\
\end{array}
\right)
\,.
\end{equation}
For an arbitrary basis $S_{11}$ can be parameterized as
\begin{equation}
S_{11}=e^{i\varphi_{11}/2}[s_{11}+s'_{11}(\hat{\mu}\cdot\vec{\sigma})]
       e^{i(\hat{\mu}\cdot \vec{\sigma})\vartheta_{11}/2}
\,,
\label{s-par}
\end{equation}
where the overall phase factor, $\varphi_{11}\equiv\theta^r_+ +\theta^r_-$, and the spin
mixing angle, $\vartheta_{11}\equiv\theta_+ -\theta_-$, are defined as the sum and difference
of the phases for the reflected spin up and spin down electrons. The two real amplitudes,
$s_{11}=(|r_+|+|r_-|)/2$ and $s'_{11}=(|r_+|-|r_-|)/2$, determine the spin filtering effect.
A similar parametrization can be carried out for each element, $S_{ij}$, of the $\mathcal{S}$
matrix.

The unitary condition, $\mathcal{S}\mathcal{S}^{\dag}=1$, combined with symmetries of
the interface, provide constraints between the values of
$\{\varphi_{ij},\vartheta_{ij},s_{ij},s'_{ij}\}$. For a specular FI interface with
inversion symmetry, the constraint of time reversal symmetry, which includes the
reversal of the ferromagnetic moment, gives $\varphi_{21}=\varphi_{11}+\pi/2$, and
$\vartheta_{21}=\vartheta_{11}$. In this case the spin mixing angle for reflection and
transmission are the same. The resulting $\mathcal{S}$ matrix simplifies, and is
conveniently expressed in the basis, $|\pm\rangle$,
\begin{eqnarray}
S_{11}=S_{22}=\;e^{i\varphi/2}\left(
\begin{array}{cc}
  \sqrt{R_{\uparrow}} e^{i\vartheta/2} & 0 \\
  0 &  \sqrt{R_{\downarrow}} e^{-i\vartheta/2}\\
\end{array}
\right) \nonumber \\
S_{21}=S_{12}=ie^{i\varphi/2}\left(
\begin{array}{cc}
  \sqrt{D_{\uparrow}} e^{i\vartheta/2} & 0 \\
  0 &  \sqrt{D_{\downarrow}} e^{-i\vartheta/2}\\
\end{array}\right)
\,,
\label{sm}
\end{eqnarray}
where $R_{\alpha}+D_{\alpha}=1$, $\alpha=\uparrow$, $\downarrow$.
The overall phase factor, $\varphi$, drops out of all observables in the
quasiclassical approximation and can be omitted. Therefore the
$\mathcal{S}$ matrix is described by three parameters: the transparencies for
spin up and spin down electrons, $D_{\uparrow}$ and $D_{\downarrow}$, and the
spin mixing angle, $\vartheta$.

If we also have reflection symmetry in a plane perpendicular to the
interface, then $S_{ij}(-\mathbf{p_{\parallel}})=S_{ij}(\mathbf{p_{\parallel}})$.
This implies that that the S matrix for hole scattering is simply
$\underline{S}_{ij}=S_{ij}$. This model of a ferromagnetic interface
defined by Eq. (\ref{sm}), as well as special cases without spin filtering, have
been discussed previously by several authors.\cite{tok88,fog00,bar02c}
The reflection and transmission probabilities, $\{R_{\alpha},D_{\alpha},\vartheta\}$
are functions of the direction of the trajectory of an incident quasiparticle, $\hat{\v{p}}_f$,
and depend on material parameters such as the band gap, $E_g$, the Fermi velocities of
the electrons in the two metallic leads, $\v{v}_{fi}$,
exchange field, $h$, barrier thickness, $l$, etc.

\begin{figure}
\includegraphics[width=3.3in]{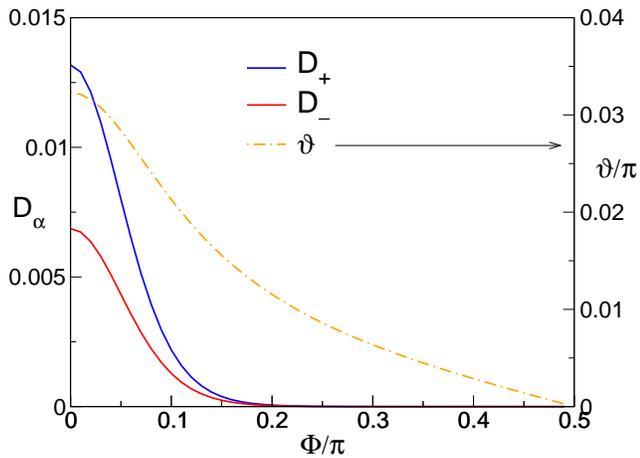}
\caption{The spin up ($D_{+}$) and spin down ($D_{-}$) transmission probabilities, and the spin mixing
         angle as a function of the angle of incidence,
         $\Phi=\arccos(\hat{\v{p}}_f\cdot\hat{z})$. The model parameters for
         the S-FI-S interface are described in the text.}
\label{fig2}
\end{figure}

To illustrate the typical parameters for spin mixing and spin filtering by a FI
interface consider a FI barrier with a band gap of $E_g=0.825$ eV and and exchange splitting
of $h=0.18$ eV, between two metallic leads.\footnote{These values correspond
to values for the ferromagnetic semiconductor, EuS.\cite{hao90}}
For conduction electrons with effective mass $m^*$ equal
to the band mass of carriers in the FI we can calculate the spin-mixing angle
and the transmission probabilities for spin up and spin down conduction electrons
at normal incidence. A barrier of width $l=0.5$ nm gives $D_{\uparrow}=0.013$,
$D_{\downarrow}=0.007$, and $\vartheta=0.032\,\pi$. The ratio $D_{\downarrow}/D_{\uparrow}$
vanishes exponentially as the barrier thickness increases, and the spin mixing angle
$\vartheta$ saturates at $0.0348\pi$, the spin mixing angle for a perfectly reflecting
FI surface. For angles away from normal incidence the effective barrier thickness
increases and the corresponding transmission probabilities decrease rapidly away from
normal incidence as shown in Fig. \ref{fig2}. The spin mixing angle also decreases with
the angle of incidence, and vanishes for grazing incidence.

The $\mathcal{S}$ matrix model in Eq. (\ref{sm}) is sufficiently general
to account for the essential features of spin-active scattering by a
clean, ferromagnetic metallic (FM) layer.\cite{bar02c} For example,
assume the transmission and reflection of electrons by the interface
is controlled by Fermi wave-vector mismatch at the S-FM interface.
Upon entering the FM layer, the Fermi momenta for majority (spin up)
and minority (spin down) electrons changes to $p_f^{\pm}=\sqrt{(E_f\pm h)2m^*}$,
respectively. As a result the transmitted majority- and minority-spin
electrons acquire a relative phase.

For sufficiently large angles of incidence the normal component of the
momentum of the spin down electrons in the FM, $p_z^{-}=\sqrt{(E_f\cos^2\Phi-h)2m^*}$,
vanishes and becomes imaginary for larger angles of incidence. Thus, these
spin down electrons can only tunnel through the FM barrier. Since the charge
and heat currents are dominated by trajectories close to normal incidence
we consider the transmission probabilities and spin-mixing angle in the small
angle limit near normal incidence, where both $p_z^{+}$ and
$p_z^{-}$ are real. If we further assume the exchange field is relatively weak,
$h\ll E_f$, then to the leading order in $h/E_f$,
\begin{eqnarray}
&&\vartheta=\theta_{+}-\theta_{-}\simeq
 (p_z^{+}-p_z^{-})l
\\
&&D_{\alpha}\simeq
1-(p_z^{\alpha}/p_z-p_z/p_z^{\alpha})^2\sin^2(p_z^{\alpha}l)/4
\,.
\end{eqnarray}
Thus, for normal incidence the spin mixing angle is of order
$\vartheta\sim p_fl\,(h/E_f)$, which can easily approach $\vartheta\to\pi$.
The electrons of both spin species are nearly perfectly transmitted,
$D_{\alpha}\simeq 1-(h/ 2E_f)^2$, and the spin filtering effect is
negligibly small,
\begin{equation}
{D_{\uparrow}-D_{\downarrow}\over D_{\uparrow}+D_{\downarrow}}
  \sim \mathcal{O}\left({h\over 2E_f}\right)^3
\,.
\end{equation}
Thus, the dominant effect of the FM interface is spin mixing, and an approximate
form for the $\mathcal{S}$ matrix of an ideal FM interface is given by
$S_{11}=S_{22}\simeq 0$ and
\begin{equation}
S_{12}=S_{21}\simeq e^{i\varphi}e^{i(\vec{\sigma}\cdot\hat{\mu})\vartheta/2}
\,.
\label{fm}
\end{equation}
Starting from Eq. (\ref{fm}) a more detailed model for the $\mathcal{S}$ matrix of
a FM interface can be constructed by adding a thin nonmagnetic
insulating layer, with transparency $D$, inside the ideal FM, which
may describe an interfacial dielectric barrier. The composite $\mathcal{S}$ matrix
of this structure takes the form of
Eq. (\ref{sm}) with $D_{\uparrow}=D_{\downarrow}=D$.

However, there presumably exist a wide variety of spin active interfaces, described by any
physically allowed value of $\vartheta$ and $D_{\alpha}$. Thus, the calculations that follow
are carried out for a broad range of values of $\vartheta$ and $D_{\alpha}$.

\section{FM \& FI point contacts}

We now illustrate the application of the boundary conditions
by calculating some representative transport properties
for both NFS and SFS point contacts. These calculations highlight the role
of the spin mixing angle in modifying the local spectrum near the point
contact and in modifying the effective transmission coefficient for
excitations that carry currents across the interface of the point contact.

Although the formalism is applicable to superconductors with any pairing symmetry,
the calculations described here are for spin-singlet, $s$-wave superconductors.
For a point contact the radius of the contact is much smaller than the
coherence length. In this limit the pairbreaking effect of the FM on the
magnitude of the order parameter can be neglected, and the voltage
drop occurs at the contact because of the large Sharvin resistance.\cite{zha03}
Then at the point contact the Riccati amplitudes
$\gamma^{R/A}_j,\tilde{\gamma}^{R/A}_j,x^K_j,\tilde{x}^K_j$ take
their local, bulk equilibrium values given by,\cite{esc00}
\begin{eqnarray}
&&\gamma^R_j(|\epsilon|<\Delta_j)=
-(i\sigma_2)e^{-i\delta_j+i\phi_j},
\label{bkv} \\
&&\gamma^R_j(|\epsilon|>\Delta_j)=
-(i\sigma_2)\mathrm{sgn}(\epsilon)e^{-\delta_j+i\phi_j},\\
&&x^K_j=(1-|\gamma_j^R|^2)\tanh {\epsilon-eV_j\over 2T_j}
\end{eqnarray}
where we introduced the dimensionless parameters,
\begin{eqnarray}
\delta_j=\arccos{\epsilon\over \Delta_j},\;|\epsilon|<\Delta_j
\label{delta1}
\\
\cosh\delta_j={|\epsilon|\over \Delta_j},\;|\epsilon|>\Delta_j
\,,
\label{delta2}
\end{eqnarray}
and $\Delta_j$ is the gap, $V_j$ is the potential, $T_j$ is the
temperature, and $\phi_j$ is the phase of superconductor on side $j=1,2$.
Application of the boundary conditions,
Eqs. (\ref{first})-(\ref{final}), is straight forward; we obtain
$\Gamma^{R/A}_j$, $\tilde{\Gamma}^{R/A}_j$, $X^K_j$ and $\tilde{X}^K_j$
from which we construct the quasiclassical Green's functions.

\subsection{NFS conductance}

Consider the electrical conductance of an NFS contact at fixed temperature, $T$,
and voltage bias, $V$. Due to both spin-mixing and the proximity effect,
the local density of states (DOS) of the superconductor deviates from the bulk
BCS form. Surface states appear below the gap, and a splitting of the DOS for
spin up ($+$) and spin down ($-$) excitations develops for any $0<\vartheta<\pi$,
\begin{eqnarray}
N_{\pm}(|\epsilon|<\Delta)=N_f { 1-R_{\uparrow}R_{\downarrow}
\over 1+ R_{\uparrow}R_{\downarrow} -2\sqrt{
R_{\uparrow}R_{\downarrow}} \cos(2\delta\mp \vartheta)}
&\,&\,
\\
N_{\pm}(|\epsilon|>\Delta)=N_f {
e^{2\delta}-e^{-2\delta}R_{\uparrow}R_{\downarrow} \over
e^{2\delta}+e^{-2\delta} R_{\uparrow}R_{\downarrow} -2\sqrt{
R_{\uparrow}R_{\downarrow}} \cos\vartheta}
&\,&\,
\end{eqnarray}
where $N_f$ is the density of states at the Fermi level, and $\delta$ is defined
in Eq. (\ref{delta1}) and Eq. (\ref{delta2}). For perfect reflection, $R_{\uparrow}=R_{\downarrow}=1$,
there is a true surface bound state, analogous to the Shiba state bound to a
magnetic impurity in an s-wave superconductor.\cite{shi68} For finite
transmission, the surface bound states broaden into resonances due to the
proximity coupling with the normal metal. In the tunnelling limit, i.e. for low
transmission with $D_{\uparrow}\approx D_{\downarrow} = D\ll 1$,
$N_{\pm}(\epsilon)$ exhibits a relatively sharp resonance peak below the gap at
$\epsilon_{\pm}\simeq\pm\Delta\cos(\vartheta/2)$ with a width of
order $\gamma\simeq D\Delta/2$.\cite{lof02a} For higher values of the transmission
probability the resonances broaden into a sub-gap continuum.

\begin{figure}
\includegraphics[width=3.3in]{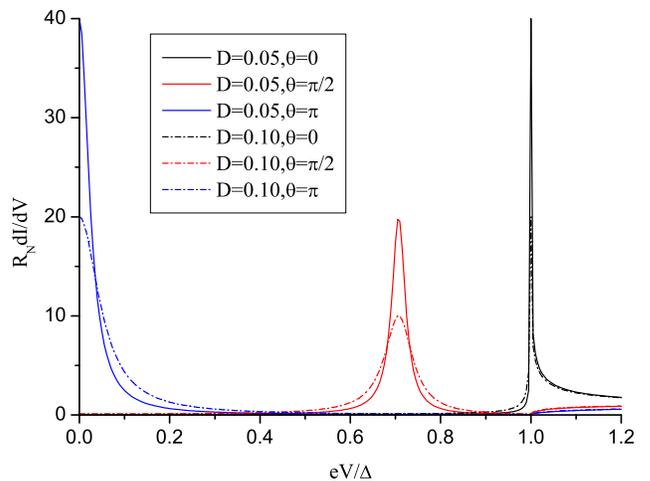}
\caption{Differential conductance, $dI/dV$, for NFS contacts with different
spin mixing angles, $\vartheta=0,\pi/2,\pi$ at $T=0$ and for
$D_{\uparrow}=D_{\downarrow}\equiv D=0.05$ and $D=0.1$.
The normal-state resistance of the point
contact is $R_N=(e^2N_f D\,\mathcal{A})^{-1}$, where $\mathcal{A}$ is the area of the
contact.}
\label{dIdV-NIS}
\end{figure}

\begin{figure}
\includegraphics[width=3.3in]{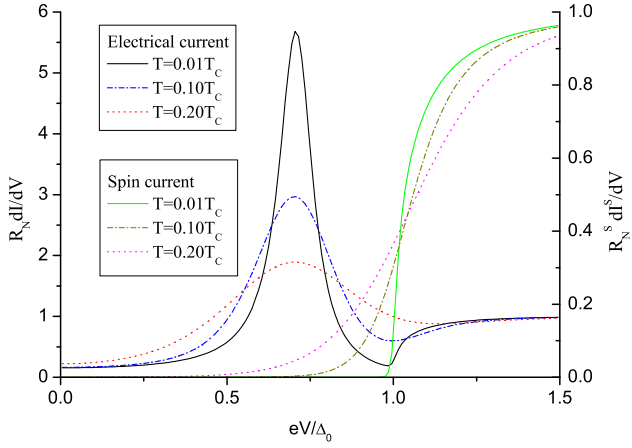}
\caption{Differential charge conductance, $dI/dV$, and spin conductance, $dI^s/dV$,
for an NFS contact with spin mixing angle, $\vartheta=\pi/2$ and transparencies,
$D_{\uparrow}=2 D_{\downarrow}=0.2$, at temperatures, $T=0.01T_c$, $T=0.1T_c$ and $T=0.2T_c$.
Note that the voltage is normalized in units of $\Delta_0$, the gap at $T=0$.
}
\label{dIdV_spin-filter-NIS}
\end{figure}

The differential conductance for low-transmission junctions reflects the resonance states which transport
charge via resonant Andreev reflection. The spectral current density, $j(\epsilon)$, can be
calculated from the solution for the quasiclassical propagators at the interface,
\begin{eqnarray}
{j^{>}\over j_N}=
{4\cosh\delta[e^{-\delta}D_{\uparrow}D_{\downarrow}/(D_{\uparrow}+D_{\downarrow})+\sinh\delta ]
            \over
e^{2\delta}+e^{-2\delta}R_{\uparrow}R_{\downarrow}-2\sqrt{R_{\uparrow}R_{\downarrow}}\cos\vartheta}
\\
{j^{<}\over j_N}=2\sum_{\pm}
{D_{\uparrow}D_{\downarrow}/ (D_{\uparrow}+D_{\downarrow}) \over
1+ R_{\uparrow}R_{\downarrow} -2\sqrt{ R_{\uparrow}R_{\downarrow}}
\cos(2\delta\pm \vartheta)}
\,,
\end{eqnarray}
where $j^{\gtrless}$ is the spectral current density for $|\epsilon|\gtrless\Delta$
and $j_N\propto e(D_{\uparrow}+D_{\downarrow})$ is the spectral current density
when the both electrodes are in the normal state. The total current density
is then given by
\begin{equation}
\hspace*{-2mm}
j=\frac{1}{2}\int d\epsilon\, j(\epsilon)[\tanh\left({\epsilon+eV\over 2T}\right)-
                                 \tanh\left({\epsilon-eV\over 2T}\right)]
\label{totalcurrent}
\,.
\end{equation}
Figure \ref{dIdV-NIS} shows the zero temperature differential conductance for
NFS junctions with different spin mixing angles.
The proximity effect is evident as a finite sub-gap conductance even for
a non-magnetic interface. The interface resonance induced by a finite spin-mixing
angle is also clearly exhibited as a broad peak in sub-gap conductance at
$eV\approx\Delta\cos(\vartheta/2)$. Note also that the width of the resonance,
$\gamma\approx D\Delta/2$, provides a spectroscopic measure of the interface
transmission probability.
However, thermal broadening dominates the width of the sub-gap resonances in
the tunneling limit, except at very low temperatures, as shown in
Fig. \ref{dIdV_spin-filter-NIS}.

Asymmetry in the transmission probabilities,
$D_{\uparrow}\ne D_{\downarrow}$, for spin up and spin down excitations also leads
to a finite spin current. The corresponding spectral current density is given by
\begin{equation}
j^{s>}= {2\left(D_{\uparrow}-D_{\downarrow}\right)\,
\sinh(2\delta)\over e^{2\delta}+e^{-2\delta}R_{\uparrow}R_{\downarrow}-2\sqrt{R_{\uparrow}R_{\downarrow}}\cos\vartheta}
\,,
\end{equation}
for $|\epsilon|>\Delta$. In contrast to the charge current the sub-gap
spin current spectral density vanishes identically ($j^{s<}=0$) because there is no resonant
Andreev reflection for spin transport. The total spin current is then given by
Eq. \ref{totalcurrent} with $j(\epsilon)\to j^{s}(\epsilon)$.
In Fig. \ref{dIdV_spin-filter-NIS} we also show the differential spin conductance for
weak spin filtering, $D_{\uparrow}=2 D_{\downarrow}=0.2$. Note the onset of the spin
conductance at $eV=\Delta$ for $T\to 0$, and the absence of Andreev resonance peaks
in the spin conductance for non-zero spin mixing.
The spin conductance is normalized by the normal-state spin conductance of a point
contact of area $\mathcal{A}$, $(R^{s}_N)^{-1}=N_f (D_{\uparrow}-D_{\downarrow})\,\mathcal{A}$.

The limit of extreme spin-filtering provided by a
half-metallic ferromagnetic metal is discussed in detail using the transfer matrix
method to incorporate spin-mixing
at the interface between a FM and a superconductor.\cite{kop04} We
obtain the same results as Ref. \onlinecite{kop04} for the current of a half-metallic
ferromagnetic-superconductor point contact by setting $D_{\uparrow}=D$ and
$D_{\downarrow}=0$. In this limit the sub-gap conductance from resonant Andreev
reflection is completely suppressed by the spin-filtering effect.

\subsection{SFS thermal conductance}

In a recent report we discussed the role of Andreev bound states in regulating
quasiparticle transport of heat through point-contact Josephson weak
links.\cite{zha03,zha04} Spin mixing generates a spin-resolved spectrum of Andreev
bound states at a spin-active point contact even in the absence of a phase bias.
We apply the formalism and boundary conditions developed in previous sections to investigate
the effect of spin mixing on the phase sensitive heat transport in temperature biased
SFS point contacts. We show that the relative phase shift of spin up and down
electrons, together with the phase bias $\phi=\phi_2-\phi_1$, determines the spectrum
of Andreev bound states at the point contact. The effects
of these states on the transmission probability of continuum excitations that transport heat is
calculated.

The thermal conductance of the point contact
is defined by the ratio of the total heat current and the temperature bias $\delta
T=T_2-T_1$ in the limit $\delta T\rightarrow 0$. The results reported below are
normalized by the normal-state thermal conductance at $T_c$,
$\kappa_N=\frac{\pi^2}{12}\Area N_f v_f T_c (D_{\uparrow}+D_{\downarrow})$,
where $\Area$ is the area of the point contact.

Following the same line of argument as described for the NFS contact we apply the
boundary conditions, Eqs. (\ref{first})-(\ref{final}), to construct the Green
functions for the SFS contact. The ABS spectrum is straightforward to calculate
and has been discussed in the context of the Josephson current of the SFS weak
link by several authors.\cite{fog00,cue01,bar02c} There are
two branches (labelled as $\pm$) of spin-up Andreev bound states with energies,
\begin{equation}
\epsilon^{\uparrow}_{\pm}=\Delta\;\mathrm{sgn}\left(\sin{\vartheta\pm\rho\over2}\right)
                          \cos{\vartheta\pm \rho\over2}
\,,
\label{absup}
\end{equation}
where the angle $\rho$ is defined as
\begin{equation}
\rho=\arccos(\sqrt{R_{\uparrow}R_{\downarrow}}+
\sqrt{D_{\uparrow}D_{\downarrow}}\cos{\phi})
\,.
\end{equation}
The spin down bound states are at energies
\begin{equation}
\epsilon^{\downarrow}_{\pm}(\vartheta)
=\epsilon^{\uparrow}_{\pm}(-\vartheta)
=-\epsilon^{\uparrow}_{\mp}(\vartheta)
\,.
\label{absdown}
\end{equation}
At $\vartheta=0$ or $\pi$, the ABS spectra is degenerate with respect to spin. For
$0<\vartheta<\pi$ the spin degeneracy is lifted, thus, the typical bound state
spectrum has four branches, two branches per spin direction for $\rho\neq 0$.
Branches with opposite spin are ``mirror reflections'' of one another with respect
to the Fermi energy ($\epsilon=0$). In the tunnelling limit, $D_{\uparrow}=D_{\downarrow}=D\rightarrow
0$, and to leading order in $\rho\ll \vartheta$, the splitting of the spin up
states is given by
\begin{equation}
\epsilon^{\uparrow}_{\pm}\simeq\Delta[\cos{\vartheta\over 2}\mp
\sin{\vartheta\over 2}\sqrt{D}\sin{\phi\over 2}]
\label{tunnelABS}
\,.
\end{equation}

The spectral weight of an ABS comes at the expense of the continuum
spectrum ($|\epsilon|>\Delta$). In addition there is asymmetry with
respect to spin, $N^{\uparrow}(|\epsilon|>\Delta)\neq
N^{\downarrow}(|\epsilon|>\Delta)$, and as a result spin up and
spin down quasiparticles contribute differently to the heat
current. To compute the heat current, we follow the procedure
described in Ref. \onlinecite{zha03}. The boundary conditions for
the distribution functions, Eqs. (\ref{third}) and (\ref{seven}),
enable us to obtain an analytical result for the Keldysh Green's
function at the point contact. From this result we can calculate
the spectral density for the heat current for the set of
trajectories, $\{1i,1o,2i,2o\}$. The spectral heat current contains
contributions from spin up and spin down electron-like and hole-like
quasiparticles,
\begin{equation}
j^h(\epsilon,\hat{\v{p}}_f) = {\epsilon \over 4\pi i}
\mathrm{Tr}[\hat{g}^K_{1i}-\hat{g}^K_{1o}]= {\epsilon\over 4\pi
i}\mathrm{Tr} [\hat{g}^K_{2o}-\hat{g}^K_{2i}]
\,,
\end{equation}
where $\mathrm{Tr}$ represents the trace of Nambu matrix propagators.
It is straightforward to show that the spectral heat current can be
expressed in an intuitive form by introducing the effective transmission
coefficient $\D$ for heat transport,
\begin{equation}
j^h(\epsilon,\hat{\v{p}}_f)=-2\epsilon\left[\tanh{\epsilon\over
2T_1}-\tanh{\epsilon\over 2T_2}\right]\,\D(\epsilon,\hat{p}_f)
\,.
\end{equation}
The contributions to $\D$ come from direct ($e\to e$) transmission and
branch conversion ($e(h)\to h(e)$) transmission channels,
$\D =\D_{e\to e}+\D_{e\to h}$,
\begin{eqnarray}
\D_{e\to e}&=&[(D_{\uparrow}+D_{\downarrow})\cosh(2\delta)
            -2\sqrt{D_{\uparrow}D_{\downarrow}}
              \cos\vartheta\cos\phi)] \nonumber \\
           &&\times \sinh^2\delta /\mathcal{Z}^2   \\
\D_{e\to h}&=&[(D_{\uparrow}+D_{\downarrow})-2D_{\uparrow}D_{\downarrow}
            -2\sqrt{R_{\uparrow}R_{\downarrow}
              D_{\uparrow}D_{\downarrow}}\cos\phi] \nonumber \\
           &&\times \sinh^2\delta /\mathcal{Z}^2  \\
             \mathcal{Z}^2&=&[\sqrt{R_{\uparrow}R_{\downarrow}}+\sqrt{D_{\uparrow}
              D_{\downarrow}}\cos{\phi}-\cosh(2\delta)\cos\vartheta]^2 \nonumber \\
           &&+\sinh^2(2\delta)\sin^2\vartheta
\,.
\end{eqnarray}
In the normal state $\D\to(D_{\uparrow}+D_{\downarrow})/2$, and spin mixing
has no effect on the quasiparticle transport. However, in the superconducting
state the transmission coefficient $\D$ for quasiparticles of energy $\epsilon$
and momentum $\v{p}_f$ is sensitive to both the phase bias, $\phi$, and the spin mixing
angle, $\vartheta$.

Consider first the case with $\phi=0$. Spin
mixing leads to bound states at
$\epsilon_\mathrm{B}=\pm\Delta\cos(\vartheta/2)$. For the case in which
$D_{\uparrow}=D_{\downarrow}=D$, only the direct transmission channel contributes,
i.e. $\D_{e\to h}(\phi=0)=0$ and
\begin{equation}
\D_{e\to e}(\phi=0) = D{\epsilon^2-\Delta^2\over
               \epsilon^2-\Delta^2\cos^2{\vartheta\over 2}}<D
\,.
\end{equation}
Thus, quasiparticle transmission is suppressed by the spin mixing effect
for any value of the normal-state transparency, $D$, and for
all energies. The suppression is most severe at $\vartheta=\pi$,
i.e. when the ABS is more strongly bound.\footnote{The situation is
analogous to a pinhole junction with phase
bias $\phi=\vartheta$, in which case the bound states are at $\pm\Delta
\cos(\phi/2)$. The only difference is that the normal-state transmission
probability, $D$, can take any value, while in the case of the phase-biased
pinhole $D=1$.}
Thus for $\phi=0$ spin mixing suppresses heat transport for any value of the normal
state transparency.

\begin{figure}
\includegraphics[width=3.3in]{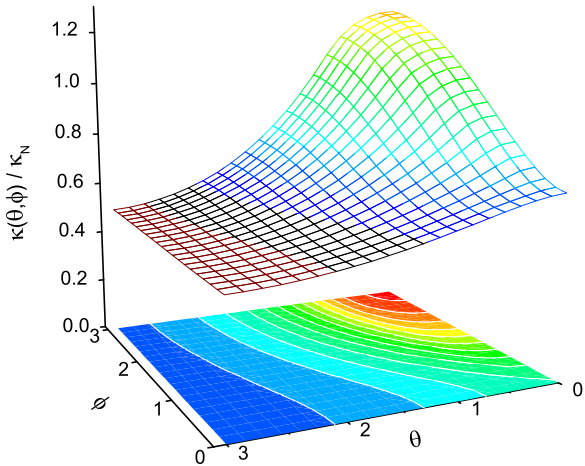}
\caption{The thermal conductance of an SFS point contact
with $D_{\uparrow}=D_{\downarrow}=0.1$ at $T=0.8T_c$. The thermal
conductance is normalized to its value at $T_c$. $\phi$ and
$\vartheta$ change from 0 to $\pi$.} \label{tunnel-heat}
\end{figure}

For a spin-inactive point contact ($\vartheta=0$) in tunnelling limit, $D\ll 1$, it is
known that $\D(\epsilon)$ has a resonance peak at
\begin{equation}
\epsilon_{\mathrm{res}}/\Delta=1+{1\over 2}D\sin^2{\phi\over
2}+\mathcal{O}(D^2)
\,,
\end{equation}
which is a reflection of a shallow bound state just below the gap edge at
$\epsilon_{\mathrm{B}}/\Delta=1-{1\over 2}D\sin^2{\phi\over 2}+\mathcal{O}(D^2)$.
Tuning $\phi$ from 0 to $\pi$ leads to an increase in the thermal conductance
because of resonant transmission of quasiparticles at
$\epsilon\approx\epsilon_{\mathrm{res}}$.\cite{zha03} Enhanced
transmission still exists for SFS point contacts, but
as $\vartheta$ increases the resonance peak of $\D(\epsilon)$
gradually vanishes. For $\vartheta=\pi$, the bound states are at energies
$\epsilon_\mathrm{B}=\pm \Delta \sqrt{D}\sin(\phi/2)$, and there is no
resonance peak of $\D(\epsilon)$. Instead $\D$ is suppressed at all energies,
\begin{equation}
\D(\vartheta=\pi)=D{\epsilon^2-\Delta^2\over
\epsilon^2-\Delta^2D\sin^2{\phi\over 2}}<D
\,.
\end{equation}
Thus, to leading order in $D$, $\D(\vartheta=\pi)\simeq
D(1-\Delta^2/\epsilon^2)$ is independent of $\phi$, so phase
modulation of the thermal conductance vanishes.
These features are shown clearly in Fig. \ref{tunnel-heat} for the thermal conductance
calculated in the tunnelling limit with $D_{\uparrow}=D_{\downarrow}=D=0.1$ and
for general values of $\vartheta$ and $\phi$.
Resonant enhancement of the conductance occurs in the vicinity of $\phi\approx\pi$
and $\vartheta\ll 1$. Increasing $\vartheta$ suppresses the overall thermal conductance
as well as the phase modulation.

\begin{figure}
\includegraphics[width=3.3in]{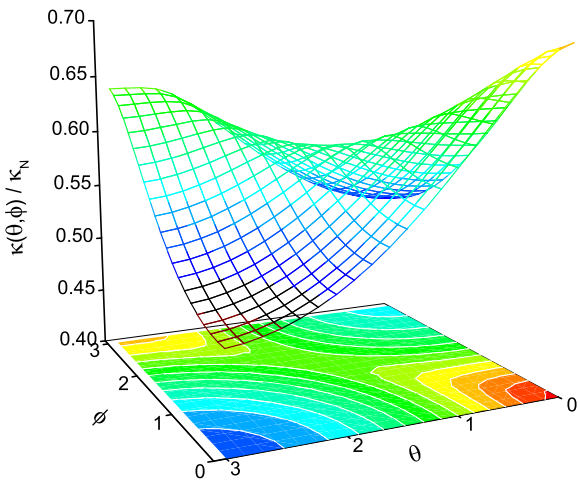}
\caption{The thermal conductance for an SFS point contact with
$D_{\uparrow}=D_{\downarrow}=0.9$ at $T=0.8T_c$.}
\label{trans-heat}
\end{figure}

Figure \ref{trans-heat} shows the thermal conductance in the high transparency limit
with $D_{\uparrow}=D_{\downarrow}=D=0.9$. At $\vartheta=0$ (a ``0'' junction),
tuning $\phi$ towards $\pi$ pushes the ABS deep into the gap, so
$\D(\epsilon)$ is increasingly suppressed from $D$, and the
thermal conductance goes down. The opposite occurs at $\vartheta=\pi$ (a ``$\pi$''
junction \cite{fog00}): tuning $\phi$ from 0 to $\pi$ pushes the ABS toward the gap edge, so the
thermal conductance increases. The phase modulation of the thermal
conductance for a general value of $\vartheta$ can be understood qualitatively
in a similar manner. The thermal conductance is maximum when the bound states are closest
to the gap edge. The different phase modulation of the thermal conductance for ``0'' junctions
(i.e. $\vartheta<\pi/2$) versus ``$\pi$'' junctions ($\vartheta>\pi/2$)
should be observable in high transparency SFS junctions; one should in principle be able
to change the spin mixing angle by varying the thickness of the FM layer, thus tuning
between ``0'' or ``$\pi$'' junction behavior. The phase of the SFS junction can then be
controlled by varying the magnetic flux linking a SQUID containing the SFS contact.

\begin{figure}
\includegraphics[width=3.3in]{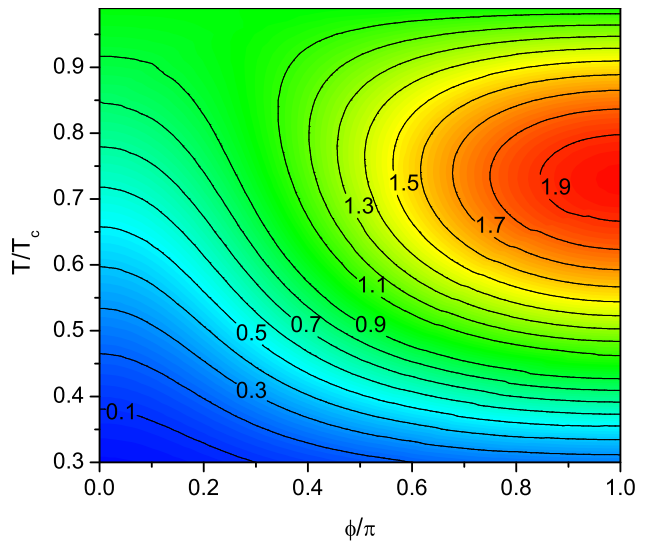}
\caption{The thermal conductance of an SFS contact in the tunnelling
limit with $D_{\uparrow}=2D_{\downarrow}=0.0013$ and $\vartheta\simeq 0.032\pi$.
}\label{contour}
\end{figure}

In contrast to FM contacts, SFS junctions with FI contacts are expected to be in the tunnelling
limit, i.e. $D_{\uparrow,\downarrow}\ll 1$. For the FI interface described in section \ref{sec_Smatrix}
with $l=0.5$ nm, $D_{\uparrow}=0.013$, $D_{\downarrow}=0.007$ and $\vartheta=0.032\pi$,
the discrimination between transparencies for different spin orientation is relatively large,
$D_{\uparrow}\simeq 2D_{\downarrow}$, but the spin mixing is weak. As a result the phase modulation of
the thermal conductance is almost the same as that of a spin-inactive tunnel junction.\cite{zha03}
Figure \ref{contour} shows a map of the thermal conductance as a function of temperature and phase
bias for the junction parameters given above. For simplicity we neglected the angular dependence
of $D_{\alpha}$ and $\vartheta$. Note that resonant transmission leads to enhancement of the
thermal conductance compared to the normal-state conductance at $T_c$ over a broad temperature
range $0.5T_c\lesssim T<T_c$ for $\phi\gtrsim \pi/2$.

\section{Discussion and Conclusion}

In this paper we derived the boundary conditions for the quasiclassical
Ricatti amplitudes, including the distribution functions, $x^K$, for general
interfaces. This completes the development of the boundary conditions for
the Riccati amplitudes at spin-active interfaces initiated
in Refs. \onlinecite{fog00}. The results summarized in section \ref{sec_boundary-conditions}
apply to a broad range of superconducting interfaces, and are specifically
applicable to non-equilibrium transport involving spin-active interfaces.
The boundary conditions can be applied to investigate dynamical properties,
are applicable for any paring symmetry. The effects of disorder are readily
described within the quasiclassical theory framework.

In the dirty limit the quasiclassical theory for conventional superconductors
can be formulated in terms of Fermi surface averaged quasiclassical Green
functions.\cite{usa70} The quasiclassical transport equations reduce to
diffusion-type (Usadel) equations. Boundary conditions for the the Fermi-surface
averaged Usadel propagators at a spin active interface has been discussed
recently in Ref. \onlinecite{her02}. A comparison between results for the
dirty SFS and NFS junctions based on the Usadel equations and boundary
conditions and the dirty limit of the Ricatti formalism with the
general spin-active boundary conditions developed here will be discussed
in a separate report.
Finally, we note that we have presented boundary conditions and results
applicable to specular interfaces. The boundary conditions can be
extended to include atomic scale roughness at a spin-active interface,
but this extension is outside the scope of this report.

The Riccati approach to the boundary problem in quasiclassical
theory is complimentary to the T-matrix approach,\cite{ser83,kie85,cue01,esc03,kop04}
which has been used to calculate the sub-harmonic structure of the current-voltage
characteristics of magnetic Josephson point contacts,\cite{and02} and recently
to study the Josephson effect and conductance of superconductors in contact with
half-metallic ferromagnets.\cite{esc03,kop04}
The advantage of the Ricatti approach is that the implementation of  the boundary condition
is given explicitly in one step, without introducing auxiliary propagators. The
Riccati formulation also makes analytical analysis more tractable. The
Riccati method has been successfully used to study the equilibrium properties,
e.g. the d.c. Josephson effect and temperature-induced $0-\pi$ transition of SFS
junctions.\cite{fog00,bar02c} As shown by Barash et al,\cite{bar02b} the Riccati
approach is especially powerful to study more complicated hybrid structures such
as SFIFS junctions.

As examples of the potential application of the newly derived boundary condition
for the non-equilibrium distribution functions, $x^K$, we
investigated the charge and spin conductances for NFS point contacts
and heat transport through temperature and phase biased SFS point contacts.
We showed how the spin-mixing angle $\vartheta$, defined as the relative phase between
spin up and down electrons upon transmission (or reflection),
controls the local excitation spectrum and the transport of
charge, spin, and energy across the point contact. Beyond this
relatively simple application, we expect that new results
and novel physics can be explored for a broad range of non-equilibrium
transport problems involving spin-active interfaces with the new boundary
conditions.

\begin{acknowledgments}
We thank Dr. M. Eschrig for stimulating discussions.
This work was supported in part by the NSF grant DMR 9972087, and STINT, the
Swedish Foundation for International Cooperation in Research and
Higher Education and the Wenner-Gren Foundations.
JAS acknowledges the support and hospitality of
the T11 Group at Los Alamos National Laboratory.
\end{acknowledgments}

\appendix*
\section{Derivation of Eqs. (\ref{first})-(\ref{final})}
In this appendix we outline how the boundary conditions for the
Riccati amplitudes, Eq. (\ref{first})-(\ref{final}), are
derived from the Millis-Rainer-Sauls boundary conditions.
Our starting point is the set of Eqs. (63)-(66) of Ref. \onlinecite{mil88}, which
in terms of Shelankov projection operators,\cite{she85}
\begin{equation}
\check{P}^{\lambda}_{\pm}=\left(
\begin{array}{cc}
\hat{P}^{\lambda,R}_{\pm} & \hat{P}^{\lambda,K}_{\pm} \\
0 & \hat{P}^{\lambda,A}_{\pm}
\end{array} \right)={1\over 2}(\check{1}\pm {\check{g}_{\lambda}\over
-i\pi}),\lambda\in \{1i, 1o, 2i, 2o\}
\end{equation}
can be written as
\begin{eqnarray}
\check{P}^{ 2o}_- (\hat{S}_{22} \check{P}^{ 2_i }_+ \hat{S}^{\dagger}_{22} + \hat{S}_{21} \check{P}^{ 1i }_- \hat{S}^{\dagger}_{21})\check{P}^{2o}_+ = 0  \label{mrs1}\,,\\
\check{P}^{ 2i }_+ (\hat{S}^{\dagger}_{22} \check{P}^{ 2o }_- \hat{S}_{22} + \hat{S}^{\dagger}_{12} \check{P}^{ 1o }_+ \hat{S}_{12}) \check{P}^{ 2i }_- = 0 \label{mrs2} \,,\\
\check{P}^{ 1i }_+ (\hat{S}^{\dagger}_{11} \check{P}^{ 1o }_+ \hat{S}_{11} + \hat{S}^{\dagger}_{21} \check{P}^{ 2o }_- \hat{S}_{21}) \check{P}^{ 1i }_- = 0 \label{mrs3} \,,\\
\check{P}^{ 1o }_- (\hat{S}_{11} \check{P}^{ 1i }_-
\hat{S}^{\dagger}_{11} + \hat{S}_{12} \check{P}^{ 2i
}_+\hat{S}^{\dagger}_{12})\check{P}^{ 1o }_+ = 0
\,.
\label{mrs4}
\end{eqnarray}
Following the convention in literature, we denote a
Keldysh$\otimes$Nambu matrix with check accent, a Nambu
(particle-hole$\otimes$spin) matrix with hat accent, and a spin
matrix without any accent. We use the \textit{ansatz} of
Eschrig\cite{esc00} to parameterize the projectors
in terms of Riccati amplitudes, e.g. for $\check{P}_+^{1i}$ we have
\begin{eqnarray}
\hat{P}^{R}_+ &=& \left(
\begin{array}{c}
1 \\
-\Gt^R_1
\end{array} \right)
(1-\gamma^R_1 \Gt^R_1)^{-1} (1,\; \gamma^R_1) \\
\hat{P}^{A}_+ &=& \left(
\begin{array}{c}
-\Gamma^A_1 \\
1
\end{array} \right)
(1-\gt^A_1 \Gamma^A_1)^{-1} (\gt^A_1,\; 1)
\\
\hat{P}^{K}_{+}&=&\hat{P}^{R}_+  \left(
\begin{array}{cc}
x^K_1 & 0 \\
0 & 0
\end{array} \right) \hat{P}^{A}_- + \hat{P}^{R}_-  \left(
\begin{array}{cc}
0 & 0 \\
0 & \Xt^K_1
\end{array} \right)\hat{P}^{A}_+
\end{eqnarray}
where we omitted the superscript $1i$. The strategy to solve Eqs. (\ref{mrs1})-(\ref{mrs4}) is
to reduce the order of the equation set by exploiting the properties of the projectors
and to decompose the equations for the Keldysh$\otimes$Nambu matrices into
equations of spin matrices.

Take the retarded components of Eqs. (\ref{mrs1})-(\ref{mrs4}) and
plug in the expressions for $\hat{P}^{\lambda,R}_{\pm}$.
Each of the four equations of Nambu matrices
collapses into an equation for the coherence functions which are
spin matrices. For example, the retarded part of Eq. (\ref{mrs2})
and (\ref{mrs3}) give
\begin{widetext}
\begin{eqnarray}
&&\beta^R_{12}(1-\Gamma^R_1\gt^R_1)^{-1}(\Gamma^R_1
\St_{12}-S_{12}\gamma^R_2) =
(\Sd_{22}\Gamma^R_2-\gamma^R_2\Std_{22})(1-\gt^R_2\Gamma^R_2)^{-1}\tilde{\beta}^R_{22}, \label{R1}\\
&&\beta^R_{11}(1-\Gamma^R_1\gt^R_1)^{-1}(\Gamma^R_1
\St_{11}-S_{11}\gamma^R_1) =
(\Sd_{21}\Gamma^R_2-\gamma^R_1\Std_{21})(1-\gt^R_2\Gamma^R_2)^{-1}\tilde{\beta}^R_{12}.
\label{R2}
\end{eqnarray}
Eqs. (\ref{R1}) and (\ref{R2}) suggest the solution
\begin{eqnarray}
(\Gamma^R_1 \St_{11}-S_{11}\gamma^R_1)(\tilde{\beta}^R_{12})^{-1}
=(\Gamma^R_1 \St_{12}-S_{12}\gamma^R_2)(\tilde{\beta}^R_{22})^{-1}, \label{GR1} \\
(\beta^R_{11})^{-1}(\Sd_{21}\Gamma^R_2-\gamma^R_1\Std_{21})
=(\beta^R_{12})^{-1}(\Sd_{22}\Gamma^R_2-\gamma^R_2\Std_{22}). \label{GR2}
\end{eqnarray}
\end{widetext}
The above results assume the inverses of $\beta^{R/A}_{ij}$ and their $\,\tilde{}$
partners exist. This is generally the case for partially transmitting interfaces. Otherwise,
Eqs. (\ref{R1}) and (\ref{R2}) can be solved rather trivially since some of the $\mathcal{S}$
matrix elements vanish. Equations (\ref{GR1})-(\ref{GR2}) immediately lead to
the result obtained by Fogelstr\"om,\cite{fog00} Eq. (\ref{alternative}), which is equivalent
to Eq. (\ref{first}), as well as
\begin{eqnarray}
\Gamma^R_2&=& \R^R_{2l}\gamma^R_2 \Std_{22} +
\T^R_{2l}\gamma^R_1 \Std_{21}  \label{G2} \\
\R^R_{2l}&=&+[(\beta^R_{12})^{-1}\Sd_{22}-(\beta^R_{11})^{-1}\Sd_{21}]^{-1}
(\beta^R_{12})^{-1} \label{G22}\\
\T^R_{2l}&=&-[(\beta^R_{12})^{-1}\Sd_{22}-(\beta^R_{11})^{-1}\Sd_{21}]^{-1}
(\beta^R_{11})^{-1}
\label{G23}
\,.
\end{eqnarray}
It is also straightforward to verify that the solution, Eq.  (\ref{alternative}) and
Eqs. (\ref{G2}-\ref{G23}), indeed satisfy
Eqs. (\ref{R1}) and (\ref{R2}). In a similar manner,
the boundary conditions for all retarded and advanced coherence
functions can be derived.

The Keldysh components of Eqs. (\ref{mrs1})-(\ref{mrs4}) can be simplified by
using the equations for retarded and advanced projectors. Then
after plugging in the expressions for all the projectors,
$\hat{P}^{\lambda,R/A/K}_{\pm}$, once again we find that the Nambu
matrix equations collapse into spin matrix equations for the
Riccati amplitudes. For example, the Keldysh components of Eqs.
(\ref{mrs2}) and (\ref{mrs3}) lead to
\begin{widetext}
\begin{eqnarray}
0=&& [ -\beta^R_{11}(1-\Gamma^R_1\gt^R_1)^{-1}S_{11}
+(\Sd_{21}\Gamma^R_2-\gamma^R_1\Std_{21})(1-\gt^R_2\Gamma^R_2)^{-1}\gt^R_2
S_{21} ]\;x^K_1 \nonumber \\
&&+x^K_1 \;[ -\Sd_{11}\gamma^A_1(1-\Gt^A_1\gamma^A_1)^{-1}(\Gt^A_1
S_{11}-\St_{11}\gt^A_1)
+\Sd_{21}(1-\gamma^A_2\Gt^A_2)^{-1}\alpha^A_{21} ] \nonumber \\
&&+\beta^R_{11}(1-\Gamma^R_1\gt^R_1)^{-1}\; X^K_1
\;(1-\gamma^A_1\Gt^A_1)^{-1}\alpha^A_{11} \nonumber \\
&&-\beta^R_{21}(1-\Gamma^R_2\gt^R_2)^{-1}\; X^K_2
\;(1-\gamma^A_2\Gt^A_2)^{-1}\alpha^A_{21} \nonumber \\
&&-(\Sd_{11}\Gamma^R_1-\gamma^R_1\Std_{11})(1-\gt^R_1\Gamma^R_1)^{-1}\;\xt^K_1\;
 (1-\Gt^A_1\gamma^A_1)^{-1}(\Gt^A_1 S_{11}-\St_{11}\gt^A_1) \nonumber \\
&&+(\Sd_{21}\Gamma^R_2-\gamma^R_1\Std_{21})(1-\gt^R_2\Gamma^R_2)^{-1}\;\xt^K_2\;
 (1-\Gt^A_2\gamma^A_2)^{-1}(\Gt^A_2 S_{21}-\St_{21}\gt^A_1), \label{XX1} \\
0=&&[ -\beta^R_{12}(1-\Gamma^R_1\gt^R_1)^{-1}S_{12}
+(\Sd_{22}\Gamma^R_2-\gamma^R_2\Std_{22})(1-\gt^R_2\Gamma^R_2)^{-1}\gt^R_2
S_{22} ]\;x^K_2 \nonumber \\
&&+x^K_2\; [ -\Sd_{12}\gamma^A_1(1-\Gt^A_1\gamma^A_1)^{-1}(\Gt^A_1
S_{12}-\St_{12}\gt^A_2)
+\Sd_{22}(1-\gamma^A_2\Gt^A_2)^{-1}\alpha^A_{22} ] \nonumber \\
&&+\beta^R_{12}(1-\Gamma^R_1\gt^R_1)^{-1}\; X^K_1
\;(1-\gamma^A_1\Gt^A_1)^{-1}\alpha^A_{12} \nonumber \\
&&-\beta^R_{22}(1-\Gamma^R_2\gt^R_2)^{-1}\; X^K_2
\;(1-\gamma^A_2\Gt^A_2)^{-1}\alpha^A_{22} \nonumber \\
&&-(\Sd_{12}\Gamma^R_1-\gamma^R_2\Std_{12})(1-\gt^R_1\Gamma^R_1)^{-1}\;\xt^K_1\;
 (1-\Gt^A_1\gamma^A_1)^{-1}(\Gt^A_1 S_{12}-\St_{12}\gt^A_2) \nonumber \\
&&+(\Sd_{22}\Gamma^R_2-\gamma^R_2\Std_{22})(1-\gt^R_2\Gamma^R_2)^{-1}\;\xt^K_2\;
 (1-\Gt^A_2\gamma^A_2)^{-1}(\Gt^A_2 S_{22}-\St_{22}\gt^A_2). \label{XX2}
\end{eqnarray}
\end{widetext}
Again, if the inverses of $\beta^{R/A}_{ij}$ do not exist
Eqs. (\ref{XX1}) and (\ref{XX2}) simplify and are readily solved.
For the general case, in order to solve for $X^K_1$ and $X^K_2$,
we construct
$(\beta^R_{21})^{-1}\times(Eq.\,\ref{XX1})\times(\alpha^A_{21})^{-1}-(\beta^R_{22})^{-1}\times(Eq.\,\ref{XX2})\times(\alpha^A_{22})^{-1}$ and
$(\beta^R_{11})^{-1}\times(Eq.\,\ref{XX1})\times(\alpha^A_{11})^{-1}-(\beta^R_{12})^{-1}\times(Eq.\,\ref{XX2})\times(\alpha^A_{12})^{-1}$ to
obtain transformed equations, Eq. (A.10)$'$ and Eq. (A.11)$'$, which are not reproduced here.
The transformed Eq. (A.10)$'$ contains only $X^K_1,x^K_1,x^K_2$ and $\xt^K_2$. We regularize each
term by carrying out a series of transformations. For example, the terms proportional to $x^K_1$ are transformed to
\begin{widetext}
\begin{eqnarray}
&+&(\beta^R_{21})^{-1}\beta^R_{11}(1-\Gamma^R_1\gt^R_1)^{-1}
(\A^R_{1l}\;\gt^R_2 S_{21}-S_{11})
\;x^K_1\;(\alpha^A_{21})^{-1} +(\beta^R_{21})^{-1}\;x^K_1\;
(\Sd_{21}\gamma^A_2\;\tilde{\A}^A_{1r} -\Sd_{11})
(1-\gamma^A_1\Gt^A_1)^{-1}\alpha^A_{11}(\alpha^A_{21})^{-1} \nonumber \\
&+&(\beta^R_{21})^{-1}\;x^K_1\;(\alpha^A_{21})^{-1}
\,,
\end{eqnarray}
and further into
\begin{eqnarray}
&-&
(\beta^R_{21})^{-1}\beta^R_{11}(1-\Gamma^R_1\gt^R_1)^{-1}
\R^R_{1l}(\Sd_{11})^{-1} x_1 (S_{11})^{-1}\tilde{\R}^A_{1r}
(1-\gamma^A_1\Gt^A_1)^{-1}\alpha^A_{11}(\alpha^A_{21})^{-1} \nonumber\\
&+&(\beta^R_{22})^{-1}\beta^R_{12}(1-\Gamma^R_1\gt^R_1)^{-1}
\R^R_{1l}(\Sd_{11})^{-1} x_1 (S_{11})^{-1}\tilde{\R}^A_{1r}
(1-\gamma^A_1\Gt^A_1)^{-1}\alpha^A_{12}(\alpha^A_{22})^{-1}
\,,
\end{eqnarray}
with the help of algebraic identities such as
\begin{eqnarray}
&&\A^R_{1l}\;\gt^R_2 S_{21}+\R^R_{1l}=S_{11},\; \A^R_{1l}\;\gt^R_2 S_{22}+\T^R_{1l}=S_{12},\\
&&1-\Gamma^R_1\gt^R_1=\R^R_{1l}\beta^R_{11}
+\T^R_{1l}\beta^R_{12},\\
&&(\beta^R_{21})^{-1}\beta^R_{11}(1-\Gamma^R_1\gt^R_1)^{-1}
\T^R_{1l}\beta^R_{12}
=(\beta^R_{22})^{-1}\beta^R_{12}(1-\Gamma^R_1\gt^R_1)^{-1}
\R^R_{1l}(\beta^R_{11})
\,.
\end{eqnarray}
As a result the transformed Eqs. (A.10)$'$ and (A.11)$'$ can be expressed as
\begin{eqnarray}
(\beta^R_{21})^{-1}\beta^R_{11}(1-\Gamma^R_1\gt^R_1)^{-1}
n_1(1-\gamma^A_1\Gt^A_1)^{-1}\beta^A_{11}(\beta^A_{21})^{-1}
=(\beta^R_{22})^{-1}\beta^R_{12}(1-\Gamma^R_1\gt^R_1)^{-1}
n_1(1-\gamma^A_1\Gt^A_1)^{-1}\beta^A_{12}(\beta^A_{22})^{-1},&&\label{fn1}\\
(\beta^R_{12})^{-1}\beta^R_{22}(1-\Gamma^R_2\gt^R_2)^{-1}
n_2(1-\gamma^A_2\Gt^A_2)^{-1}\beta^A_{22}(\beta^A_{21})^{-1}
=(\beta^R_{11})^{-1}\beta^R_{21}(1-\Gamma^R_2\gt^R_2)^{-1}
n_2(1-\gamma^A_2\Gt^A_2)^{-1}\beta^A_{21}(\beta^A_{11})^{-1},&&\label{fn2}
\end{eqnarray}
\end{widetext}
with
\begin{eqnarray}
n_1 \equiv X^K_1 - \R^R_{1l}x^K_1 \tilde{\R}^A_{1r} -
\T^R_{1l}x^K_2
\tilde{\T}^A_{1r} + \A^R_{1l} \xt^K_2 \tilde{\A}^A_{1r}
\,,\\
n_2 \equiv X^K_2 - \R^R_{2l}x^K_2 \tilde{\R}^A_{2r} -
\T^R_{2l}x^K_1 \tilde{\T}^A_{2r} + \A^R_{2l} \xt^K_1
\tilde{\A}^A_{2r}\,.
\end{eqnarray}
Obviously $n_1=n_2=0$ satisfies Eq. (\ref{fn1}) and
(\ref{fn2}), or equivalently, the original Eqs. (\ref{XX1})
and (\ref{XX2}). The equation $n_1=0$ yields the boundary
condition for $X^K_1$ in Eq. (\ref{third}). The boundary
condition for $\tilde{X}^K_j$ is derived in a similar manner
starting from Eqs. (\ref{mrs1}) and (\ref{mrs4}).

\newpage


\begin{thebibliography}{40}
\expandafter\ifx\csname natexlab\endcsname\relax\def\natexlab#1{#1}\fi
\expandafter\ifx\csname bibnamefont\endcsname\relax
  \def\bibnamefont#1{#1}\fi
\expandafter\ifx\csname bibfnamefont\endcsname\relax
  \def\bibfnamefont#1{#1}\fi
\expandafter\ifx\csname citenamefont\endcsname\relax
  \def\citenamefont#1{#1}\fi
\expandafter\ifx\csname url\endcsname\relax
  \def\url#1{\texttt{#1}}\fi
\expandafter\ifx\csname urlprefix\endcsname\relax\def\urlprefix{URL }\fi
\providecommand{\bibinfo}[2]{#2}
\providecommand{\eprint}[2][]{\url{#2}}

\bibitem[{\citenamefont{Wolf et~al.}(2001)\citenamefont{Wolf, Awschalom,
  Buhrman, Daughton, von Moln\'ar, Roukes, Chtchelkanova, and Treger}}]{wol01}
\bibinfo{author}{\bibfnamefont{S.~A.} \bibnamefont{Wolf}},
  \bibinfo{author}{\bibfnamefont{D.~D.} \bibnamefont{Awschalom}},
  \bibinfo{author}{\bibfnamefont{R.~A.} \bibnamefont{Buhrman}},
  \bibinfo{author}{\bibfnamefont{J.~M.} \bibnamefont{Daughton}},
  \bibinfo{author}{\bibfnamefont{S.}~\bibnamefont{von Moln\'ar}},
  \bibinfo{author}{\bibfnamefont{M.~L.} \bibnamefont{Roukes}},
  \bibinfo{author}{\bibfnamefont{A.~Y.} \bibnamefont{Chtchelkanova}},
  \bibnamefont{and} \bibinfo{author}{\bibfnamefont{D.~M.}
  \bibnamefont{Treger}}, \bibinfo{journal}{Science}
  \textbf{\bibinfo{volume}{294}}, \bibinfo{pages}{1488} (\bibinfo{year}{2001}).

\bibitem[{\citenamefont{Tedrow and Meservey}(1994)}]{ted94}
\bibinfo{author}{\bibfnamefont{P.~M.} \bibnamefont{Tedrow}} \bibnamefont{and}
  \bibinfo{author}{\bibfnamefont{R.}~\bibnamefont{Meservey}},
  \bibinfo{journal}{Phys. Rep.} \textbf{\bibinfo{volume}{238}},
  \bibinfo{pages}{173} (\bibinfo{year}{1994}).

\bibitem[{\citenamefont{Soulen et~al.}(1998)\citenamefont{Soulen, Byers,
  Osofsky, Nadgorny, Ambrose, Cheng, Broussard, Tanaka, Nowak, Moodera
  et~al.}}]{sou98}
\bibinfo{author}{\bibfnamefont{R.~J.} \bibnamefont{Soulen}},
  \bibinfo{author}{\bibfnamefont{J.~M.} \bibnamefont{Byers}},
  \bibinfo{author}{\bibfnamefont{M.~S.} \bibnamefont{Osofsky}},
  \bibinfo{author}{\bibfnamefont{B.}~\bibnamefont{Nadgorny}},
  \bibinfo{author}{\bibfnamefont{T.}~\bibnamefont{Ambrose}},
  \bibinfo{author}{\bibfnamefont{S.~F.} \bibnamefont{Cheng}},
  \bibinfo{author}{\bibfnamefont{P.~R.} \bibnamefont{Broussard}},
  \bibinfo{author}{\bibfnamefont{C.~T.} \bibnamefont{Tanaka}},
  \bibinfo{author}{\bibfnamefont{J.}~\bibnamefont{Nowak}},
  \bibinfo{author}{\bibfnamefont{J.~S.} \bibnamefont{Moodera}},
  \bibnamefont{et~al.}, \bibinfo{journal}{Science}
  \textbf{\bibinfo{volume}{282}}, \bibinfo{pages}{85} (\bibinfo{year}{1998}).

\bibitem[{\citenamefont{Buzdin et~al.}(1982)\citenamefont{Buzdin, Bulaevski,
  and Panyukov}}]{buz82}
\bibinfo{author}{\bibfnamefont{A.}~\bibnamefont{Buzdin}},
  \bibinfo{author}{\bibfnamefont{L.}~\bibnamefont{Bulaevski}},
  \bibnamefont{and} \bibinfo{author}{\bibfnamefont{S.}~\bibnamefont{Panyukov}},
  \textbf{\bibinfo{volume}{35}}, \bibinfo{pages}{147} (\bibinfo{year}{1982}),
  \bibinfo{note}{[JETP Lett. 35, 178 (1982)]}.

\bibitem[{\citenamefont{Buzdin and Kupriyanov}(1991)}]{buz91}
\bibinfo{author}{\bibfnamefont{A.}~\bibnamefont{Buzdin}} \bibnamefont{and}
  \bibinfo{author}{\bibfnamefont{M.~Y.} \bibnamefont{Kupriyanov}},
  \textbf{\bibinfo{volume}{53}}, \bibinfo{pages}{308} (\bibinfo{year}{1991}),
  \bibinfo{note}{[JETP Lett. 53, 321 (1991)]}.

\bibitem[{\citenamefont{Ryazanov et~al.}(2001)\citenamefont{Ryazanov, Oboznov,
  Rusanov, Veretennikov, Golubov, and Aarts}}]{rya01}
\bibinfo{author}{\bibfnamefont{V.}~\bibnamefont{Ryazanov}},
  \bibinfo{author}{\bibfnamefont{V.}~\bibnamefont{Oboznov}},
  \bibinfo{author}{\bibfnamefont{A.}~\bibnamefont{Rusanov}},
  \bibinfo{author}{\bibfnamefont{A.}~\bibnamefont{Veretennikov}},
  \bibinfo{author}{\bibfnamefont{A.}~\bibnamefont{Golubov}}, \bibnamefont{and}
  \bibinfo{author}{\bibfnamefont{J.}~\bibnamefont{Aarts}},
  \bibinfo{journal}{Phys. Rev. Lett.} \textbf{\bibinfo{volume}{86}},
  \bibinfo{pages}{2427} (\bibinfo{year}{2001}).

\bibitem[{\citenamefont{Kontos et~al.}(2002)\citenamefont{Kontos, Aprili,
  Lesueur, Genet, Stephanidis, and Boursier}}]{kon02}
\bibinfo{author}{\bibfnamefont{T.}~\bibnamefont{Kontos}},
  \bibinfo{author}{\bibfnamefont{M.}~\bibnamefont{Aprili}},
  \bibinfo{author}{\bibfnamefont{J.}~\bibnamefont{Lesueur}},
  \bibinfo{author}{\bibfnamefont{F.}~\bibnamefont{Genet}},
  \bibinfo{author}{\bibfnamefont{B.}~\bibnamefont{Stephanidis}},
  \bibnamefont{and} \bibinfo{author}{\bibfnamefont{R.}~\bibnamefont{Boursier}},
  \bibinfo{journal}{Phys. Rev. Lett.} \textbf{\bibinfo{volume}{89}},
  \bibinfo{pages}{137007} (\bibinfo{year}{2002}).

\bibitem[{\citenamefont{Guichard et~al.}(2003)\citenamefont{Guichard, Aprili,
  Bourgeois, Kontos, Lesueur, and Gandit}}]{gui03}
\bibinfo{author}{\bibfnamefont{W.}~\bibnamefont{Guichard}},
  \bibinfo{author}{\bibfnamefont{M.}~\bibnamefont{Aprili}},
  \bibinfo{author}{\bibfnamefont{O.}~\bibnamefont{Bourgeois}},
  \bibinfo{author}{\bibfnamefont{T.}~\bibnamefont{Kontos}},
  \bibinfo{author}{\bibfnamefont{J.}~\bibnamefont{Lesueur}}, \bibnamefont{and}
  \bibinfo{author}{\bibfnamefont{P.}~\bibnamefont{Gandit}},
  \bibinfo{journal}{Phys. Rev. Lett.} \textbf{\bibinfo{volume}{90}},
  \bibinfo{pages}{167001} (\bibinfo{year}{2003}).

\bibitem[{\citenamefont{{M. Fogelstr\"om}}(2000)}]{fog00}
\bibinfo{author}{\bibnamefont{{M. Fogelstr\"om}}}, \bibinfo{journal}{Phys. Rev.
  B} \textbf{\bibinfo{volume}{62}}, \bibinfo{pages}{11812}
  (\bibinfo{year}{2000}).

\bibitem[{\citenamefont{Barash and Bobkova}(2002)}]{bar02c}
\bibinfo{author}{\bibfnamefont{Y.~S.} \bibnamefont{Barash}} \bibnamefont{and}
  \bibinfo{author}{\bibfnamefont{I.~V.} \bibnamefont{Bobkova}},
  \bibinfo{journal}{Phys. Rev. B} \textbf{\bibinfo{volume}{65}},
  \bibinfo{pages}{144502} (\bibinfo{year}{2002}).

\bibitem[{\citenamefont{Koshina and Krivoruchko}(2001)}]{kos01}
\bibinfo{author}{\bibfnamefont{E.}~\bibnamefont{Koshina}} \bibnamefont{and}
  \bibinfo{author}{\bibfnamefont{V.}~\bibnamefont{Krivoruchko}},
  \bibinfo{journal}{Phys. Rev. B} \textbf{\bibinfo{volume}{63}},
  \bibinfo{pages}{224515} (\bibinfo{year}{2001}).

\bibitem[{\citenamefont{Bergeret et~al.}(2001)\citenamefont{Bergeret, Volkov,
  and Efetov}}]{ber01}
\bibinfo{author}{\bibfnamefont{F.~S.} \bibnamefont{Bergeret}},
  \bibinfo{author}{\bibfnamefont{A.~F.} \bibnamefont{Volkov}},
  \bibnamefont{and} \bibinfo{author}{\bibfnamefont{K.~B.}
  \bibnamefont{Efetov}}, \bibinfo{journal}{Phys. Rev. Lett.}
  \textbf{\bibinfo{volume}{86}}, \bibinfo{pages}{3140} (\bibinfo{year}{2001}).

\bibitem[{\citenamefont{Golubov et~al.}(2002)\citenamefont{Golubov, Kupriyanov,
  and Fominov}}]{gol02}
\bibinfo{author}{\bibfnamefont{A.~A.} \bibnamefont{Golubov}},
  \bibinfo{author}{\bibfnamefont{M.~Y.} \bibnamefont{Kupriyanov}},
  \bibnamefont{and} \bibinfo{author}{\bibfnamefont{Y.~V.}
  \bibnamefont{Fominov}}, \bibinfo{journal}{Sov. Phys. JETP Lett.}
  \textbf{\bibinfo{volume}{75}}, \bibinfo{pages}{190} (\bibinfo{year}{2002}).

\bibitem[{\citenamefont{Barash et~al.}(2002)\citenamefont{Barash, Bobkova, and
  Kopp}}]{bar02b}
\bibinfo{author}{\bibfnamefont{Y.~S.} \bibnamefont{Barash}},
  \bibinfo{author}{\bibfnamefont{I.~V.} \bibnamefont{Bobkova}},
  \bibnamefont{and} \bibinfo{author}{\bibfnamefont{T.}~\bibnamefont{Kopp}},
  \bibinfo{journal}{Phys. Rev. B} \textbf{\bibinfo{volume}{66}},
  \bibinfo{pages}{140503 (R)} (\bibinfo{year}{2002}).

\bibitem[{\citenamefont{Blatter et~al.}(2001)\citenamefont{Blatter,
  Geshkenbein, and Ioffe}}]{bla01}
\bibinfo{author}{\bibfnamefont{G.}~\bibnamefont{Blatter}},
  \bibinfo{author}{\bibfnamefont{V.~B.} \bibnamefont{Geshkenbein}},
  \bibnamefont{and} \bibinfo{author}{\bibfnamefont{L.~B.} \bibnamefont{Ioffe}},
  \bibinfo{journal}{Phys. Rev. B} \textbf{\bibinfo{volume}{63}},
  \bibinfo{pages}{174511} (\bibinfo{year}{2001}).

\bibitem[{\citenamefont{Andersson et~al.}(2002)\citenamefont{Andersson, Cuevas,
  and Fogelstr\"om}}]{and02}
\bibinfo{author}{\bibfnamefont{M.}~\bibnamefont{Andersson}},
  \bibinfo{author}{\bibfnamefont{J.~C.} \bibnamefont{Cuevas}},
  \bibnamefont{and}
  \bibinfo{author}{\bibfnamefont{M.}~\bibnamefont{Fogelstr\"om}},
  \bibinfo{journal}{Physica C} \textbf{\bibinfo{volume}{367}}
  (\bibinfo{year}{2002}).

\bibitem[{\citenamefont{Eliashberg}(1972)}]{eli72}
\bibinfo{author}{\bibfnamefont{G.~M.} \bibnamefont{Eliashberg}},
  \bibinfo{journal}{Sov. Phys. JETP} \textbf{\bibinfo{volume}{34}},
  \bibinfo{pages}{668} (\bibinfo{year}{1972}).

\bibitem[{\citenamefont{Larkin and Ovchinnikov}(1976)}]{lar76}
\bibinfo{author}{\bibfnamefont{A.}~\bibnamefont{Larkin}} \bibnamefont{and}
  \bibinfo{author}{\bibfnamefont{Y.}~\bibnamefont{Ovchinnikov}},
  \bibinfo{journal}{Sov. Phys. JETP} \textbf{\bibinfo{volume}{41}},
  \bibinfo{pages}{960} (\bibinfo{year}{1976}).

\bibitem[{\citenamefont{Serene and Rainer}(1983)}]{ser83}
\bibinfo{author}{\bibfnamefont{J.~W.} \bibnamefont{Serene}} \bibnamefont{and}
  \bibinfo{author}{\bibfnamefont{D.}~\bibnamefont{Rainer}},
  \bibinfo{journal}{Phys. Rep.} \textbf{\bibinfo{volume}{101}},
  \bibinfo{pages}{221} (\bibinfo{year}{1983}).

\bibitem[{\citenamefont{Rammer and Smith}(1986)}]{ram86}
\bibinfo{author}{\bibfnamefont{J.}~\bibnamefont{Rammer}} \bibnamefont{and}
  \bibinfo{author}{\bibfnamefont{H.}~\bibnamefont{Smith}},
  \bibinfo{journal}{Rev. Mod. Phys.} \textbf{\bibinfo{volume}{58}},
  \bibinfo{pages}{323} (\bibinfo{year}{1986}).

\bibitem[{\citenamefont{Kieselmann}(1985)}]{kie85}
\bibinfo{author}{\bibfnamefont{G.}~\bibnamefont{Kieselmann}}, Ph.D. thesis,
  \bibinfo{school}{Bayreuth Universit\"at} (\bibinfo{year}{1985}).

\bibitem[{\citenamefont{Cuevas and Fogelstr\"om}(2001)}]{cue01}
\bibinfo{author}{\bibfnamefont{I.~C.} \bibnamefont{Cuevas}} \bibnamefont{and}
  \bibinfo{author}{\bibfnamefont{M.}~\bibnamefont{Fogelstr\"om}},
  \bibinfo{journal}{Phys. Rev. B} \textbf{\bibinfo{volume}{64}},
  \bibinfo{pages}{104502} (\bibinfo{year}{2001}).

\bibitem[{\citenamefont{Eschrig et~al.}(2003)\citenamefont{Eschrig, Kopu,
  Cuevas, and Sch\"on}}]{esc03}
\bibinfo{author}{\bibfnamefont{M.}~\bibnamefont{Eschrig}},
  \bibinfo{author}{\bibfnamefont{J.}~\bibnamefont{Kopu}},
  \bibinfo{author}{\bibfnamefont{J.~C.} \bibnamefont{Cuevas}},
  \bibnamefont{and} \bibinfo{author}{\bibfnamefont{G.}~\bibnamefont{Sch\"on}},
  \bibinfo{journal}{Phys. Rev. Lett.} \textbf{\bibinfo{volume}{90}},
  \bibinfo{pages}{137003} (\bibinfo{year}{2003}).

\bibitem[{\citenamefont{Kopu et~al.}(2004)\citenamefont{Kopu, Eschrig, Cuevas,
  and Fogelstr\"om}}]{kop04}
\bibinfo{author}{\bibfnamefont{J.}~\bibnamefont{Kopu}},
  \bibinfo{author}{\bibfnamefont{M.}~\bibnamefont{Eschrig}},
  \bibinfo{author}{\bibfnamefont{J.~C.} \bibnamefont{Cuevas}},
  \bibnamefont{and}
  \bibinfo{author}{\bibfnamefont{M.}~\bibnamefont{Fogelstr\"om}},
  \bibinfo{journal}{Phys. Rev. B} \textbf{\bibinfo{volume}{69}},
  \bibinfo{pages}{094501} (\bibinfo{year}{2004}).

\bibitem[{\citenamefont{Za\u{i}tsev}(1984)}]{zai84}
\bibinfo{author}{\bibfnamefont{A.~V.} \bibnamefont{Za\u{i}tsev}},
  \bibinfo{journal}{Sov. Phys. JETP} \textbf{\bibinfo{volume}{59}},
  \bibinfo{pages}{1015} (\bibinfo{year}{1984}), \bibinfo{note}{[Zh. Eskp. Teor.
  Fiz. 59, 1015 (1984)]}.

\bibitem[{\citenamefont{Millis et~al.}(1988)\citenamefont{Millis, Rainer, and
  Sauls}}]{mil88}
\bibinfo{author}{\bibfnamefont{A.}~\bibnamefont{Millis}},
  \bibinfo{author}{\bibfnamefont{D.}~\bibnamefont{Rainer}}, \bibnamefont{and}
  \bibinfo{author}{\bibfnamefont{J.~A.} \bibnamefont{Sauls}},
  \bibinfo{journal}{Phys. Rev.} \textbf{\bibinfo{volume}{B38}},
  \bibinfo{pages}{4504} (\bibinfo{year}{1988}).

\bibitem[{\citenamefont{Eschrig}(2000)}]{esc00}
\bibinfo{author}{\bibfnamefont{M.}~\bibnamefont{Eschrig}},
  \bibinfo{journal}{Phys. Rev. B} \textbf{\bibinfo{volume}{61}},
  \bibinfo{pages}{9061} (\bibinfo{year}{2000}).

\bibitem[{\citenamefont{Schopohl and Maki}(1995)}]{sch95}
\bibinfo{author}{\bibfnamefont{N.}~\bibnamefont{Schopohl}} \bibnamefont{and}
  \bibinfo{author}{\bibfnamefont{K.}~\bibnamefont{Maki}},
  \bibinfo{journal}{Physica} \textbf{\bibinfo{volume}{B 204}},
  \bibinfo{pages}{214} (\bibinfo{year}{1995}).

\bibitem[{\citenamefont{Schopohl}(1996)}]{sch96}
\bibinfo{author}{\bibfnamefont{N.}~\bibnamefont{Schopohl}}, in
  \emph{\bibinfo{booktitle}{{\sl Quasiclassical Methods in Superconductivity
  and Superfluidity}}}, edited by
  \bibinfo{editor}{\bibfnamefont{D.}~\bibnamefont{Rainer}} \bibnamefont{and}
  \bibinfo{editor}{\bibfnamefont{J.~A.} \bibnamefont{Sauls}}
  (\bibinfo{year}{1996}), pp. \bibinfo{pages}{88--103},
  \bibinfo{note}{http://lanl.arxiv.org/abs/cond-mat/9804064}.

\bibitem[{\citenamefont{Eschrig et~al.}(1999)\citenamefont{Eschrig, Sauls, and
  Rainer}}]{esc99}
\bibinfo{author}{\bibfnamefont{M.}~\bibnamefont{Eschrig}},
  \bibinfo{author}{\bibfnamefont{J.~A.} \bibnamefont{Sauls}}, \bibnamefont{and}
  \bibinfo{author}{\bibfnamefont{D.}~\bibnamefont{Rainer}},
  \bibinfo{journal}{Phys. Rev. B} \textbf{\bibinfo{volume}{60}},
  \bibinfo{pages}{10447} (\bibinfo{year}{1999}).

\bibitem[{\citenamefont{Shelankov and Ozana}(2000)}]{she00}
\bibinfo{author}{\bibfnamefont{A.}~\bibnamefont{Shelankov}} \bibnamefont{and}
  \bibinfo{author}{\bibfnamefont{M.}~\bibnamefont{Ozana}},
  \bibinfo{journal}{Phys. Rev. B} \textbf{\bibinfo{volume}{61}},
  \bibinfo{pages}{7077} (\bibinfo{year}{2000}).

\bibitem[{\citenamefont{Tokuyasu et~al.}(1988)\citenamefont{Tokuyasu, Sauls,
  and Rainer}}]{tok88}
\bibinfo{author}{\bibfnamefont{T.}~\bibnamefont{Tokuyasu}},
  \bibinfo{author}{\bibfnamefont{J.~A.}~\bibnamefont{Sauls}}, \bibnamefont{and}
  \bibinfo{author}{\bibfnamefont{D.}~\bibnamefont{Rainer}},
  \bibinfo{journal}{Phys. Rev.} \textbf{\bibinfo{volume}{B38}},
  \bibinfo{pages}{8823} (\bibinfo{year}{1988}).

\bibitem[{\citenamefont{Zhao et~al.}(2003)\citenamefont{Zhao, L\"ofwander, and
  Sauls}}]{zha03}
\bibinfo{author}{\bibfnamefont{E.}~\bibnamefont{Zhao}},
  \bibinfo{author}{\bibfnamefont{T.}~\bibnamefont{L\"ofwander}},
  \bibnamefont{and} \bibinfo{author}{\bibfnamefont{J.~A.} \bibnamefont{Sauls}},
  \bibinfo{journal}{Phys. Rev. Lett.} \textbf{\bibinfo{volume}{91}},
  \bibinfo{pages}{077003} (\bibinfo{year}{2003}).

\bibitem[{\citenamefont{Shiba}(1968)}]{shi68}
\bibinfo{author}{\bibfnamefont{H.}~\bibnamefont{Shiba}},
  \bibinfo{journal}{Prog. Theor. Phys.} \textbf{\bibinfo{volume}{40}},
  \bibinfo{pages}{435} (\bibinfo{year}{1968}).

\bibitem[{\citenamefont{L{\"o}fwander et~al.}(2002)\citenamefont{L{\"o}fwander,
  Shumeiko, and Wendin}}]{lof02a}
\bibinfo{author}{\bibfnamefont{T.}~\bibnamefont{L{\"o}fwander}},
  \bibinfo{author}{\bibfnamefont{V.~S.} \bibnamefont{Shumeiko}},
  \bibnamefont{and} \bibinfo{author}{\bibfnamefont{G.}~\bibnamefont{Wendin}},
  \bibinfo{journal}{Physica C} \textbf{\bibinfo{volume}{367}},
  \bibinfo{pages}{86} (\bibinfo{year}{2002}).

\bibitem[{\citenamefont{Zhao et~al.}(2004)\citenamefont{Zhao, L\"ofwander, and
  Sauls}}]{zha04}
\bibinfo{author}{\bibfnamefont{E.}~\bibnamefont{Zhao}},
  \bibinfo{author}{\bibfnamefont{T.}~\bibnamefont{L\"ofwander}},
  \bibnamefont{and} \bibinfo{author}{\bibfnamefont{J.~A.} \bibnamefont{Sauls}},
  \bibinfo{journal}{Phys. Rev. B} \textbf{\bibinfo{volume}{69}},
  \bibinfo{pages}{14} (\bibinfo{year}{2004}).

\bibitem[{\citenamefont{Usadel}(1970)}]{usa70}
\bibinfo{author}{\bibfnamefont{K.}~\bibnamefont{Usadel}},
  \bibinfo{journal}{Phys. Rev. Lett.} \textbf{\bibinfo{volume}{25}},
  \bibinfo{pages}{507} (\bibinfo{year}{1970}).

\bibitem[{\citenamefont{Huertas-Hernando
  et~al.}(2002)\citenamefont{Huertas-Hernando, Nazarov, and Belzig}}]{her02}
\bibinfo{author}{\bibfnamefont{D.}~\bibnamefont{Huertas-Hernando}},
  \bibinfo{author}{\bibfnamefont{Y.~V.} \bibnamefont{Nazarov}},
  \bibnamefont{and} \bibinfo{author}{\bibfnamefont{W.}~\bibnamefont{Belzig}},
  \bibinfo{journal}{Phys. Rev. Lett.} \textbf{\bibinfo{volume}{88}},
  \bibinfo{pages}{047003} (\bibinfo{year}{2002}).

\bibitem[{\citenamefont{Shelankov}(1985)}]{she85}
\bibinfo{author}{\bibfnamefont{A.}~\bibnamefont{Shelankov}},
  \bibinfo{journal}{J. Low Temp. Phys.} \textbf{\bibinfo{volume}{60}},
  \bibinfo{pages}{29} (\bibinfo{year}{1985}).

\bibitem[{\citenamefont{Hao et~al.}(1990)\citenamefont{Hao, Moodera, and
  Meservey}}]{hao90}
\bibinfo{author}{\bibfnamefont{X.}~\bibnamefont{Hao}},
  \bibinfo{author}{\bibfnamefont{J.~S.} \bibnamefont{Moodera}},
  \bibnamefont{and} \bibinfo{author}{\bibfnamefont{R.}~\bibnamefont{Meservey}},
  \bibinfo{journal}{Phys. Rev. B} \textbf{\bibinfo{volume}{42}},
  \bibinfo{pages}{29} (\bibinfo{year}{1990}).

\end{thebibliography}

\end{document}